\documentclass[12pt]{article}
\usepackage{amssymb}
\usepackage{latexsym}
\usepackage{graphicx,a4}
\usepackage[dvips]{epsfig}
\usepackage{psfrag}
\usepackage{axodraw}
\usepackage{xspace}

\newcommand{\nubar}{{\bar{\nu}}}
\newcommand{\slache}[1]{#1\!\!\!/}
\newcommand{\CnuB}{C\(\nu\)B\xspace}
\newcommand{\UHEnu}{UHE\(\nu\)\xspace}
\newcommand{\im}{\mathop{\mathrm{Im}}}

\begin{document}

\title{UHE neutrino damping in a thermal gas of relic neutrinos}
\author{J. C. D'Olivo, L. Nellen, S. Sahu and V. Van Elewyck\thanks{E-mail: 
vero@nucleares.unam.mx} \\
{\em \small Instituto de Ciencias Nucleares,} \\
{\em \small Universidad Nacional
  Aut\'onoma de M\'exico} \\ 
  {\em \small Circuito Exterior S/N, C.U., 04510 M\'exico D.F., Mexico}}
\date{November 2005}
\maketitle  

\begin{abstract}
We present a calculation of the damping of an
ultra-energetic cosmic neutrino (\UHEnu) travelling through the
thermal gas of relic neutrinos, using the formalism of
finite-temperature field theory. From the self-energy diagram due
to Z exchange, we obtain the annihilation cross section for an \UHEnu
interacting with an 
antineutrino from the background.
This method allows us to derive the full expression for the \UHEnu
transmission probability, 
taking into account the
momentum of relic neutrinos. 
We discuss the effect of
thermal motion on the shape of the absorption dips for different
\UHEnu fluxes as well as in the context of relic neutrino
clustering. We find that for ratios of the neutrino mass to the
relic background temperature $10^2$ or smaller, the thermal
broadening of the absorption lines could significantly affect the
determination of the neutrino mass and of the characteristics of
the population of \UHEnu sources.
\end{abstract}

\section{Introduction}
One of the predictions of Big Bang Cosmology is that the Universe
is filled with a background of neutrinos, analogous to the cosmic
microwave background, but with a lower temperature $T_{\nu 0}
\approx 1.95$ K ($1.69 \times 10^{-4}$ eV) and a number density $n_{\nu
0} \approx 56 \ \mathrm{cm}^{-3}$ per species~\cite{peebles,kolb}. The
direct detection of this cosmological relic background is
extremely difficult because of the very small interaction
cross-section of low energy neutrinos. 

It is therefore interesting
to explore the possibility of probing the cosmic neutrino
background (\CnuB) with ultra high energy neutrinos (\UHEnu). At
high energies, the \(Z\)~resonance in the \(s\)~channel for the
process \(\nu\nubar \to X\) enhances the probability for the
interaction of an
\UHEnu with the \CnuB~\cite{weiler,weiler84,gelmini,roulet93,yoshida1,yoshida2}). 
This process has also been proposed as a possible mechanism for generating UHE cosmic rays through 
the hadronic decay of the Z boson (``Z-burst''
mechanism~\cite{fargion99,weiler99,waxman98,yoshida3,zas00}). In that
context, it has been pointed out that some features of the \CnuB (in particular the relic neutrino 
masses, energy spectra and densities) might be indirectly inferred from the observed spectrum of 
UHE cosmic
rays~\cite{pas,fargion01}. 

On the other hand, the resonant production of Z through $\nu\nubar$ annihilation also results in 
absorption dips in the \UHEnu spectrum. If these absorption lines can be observed at Earth with 
the appropriated resolution, their study would lead us more directly to the same goal: to perform 
relic neutrino spectroscopy, thereby providing us with evidence for the existence of the \CnuB and 
with an independent way of determining the absolute neutrino
mass (for recent discussions on the subject, see \cite{ringwaldnew,quigg05}).

Most of the work in the literature assumes that the relic neutrinos
are at rest. For small neutrino masses, though, the average
momentum can be comparable to the neutrino mass. 
In this paper,
we compute the dominant contribution to the interaction of an
\UHEnu with the \CnuB using finite temperature field theory
(FTFT). 
This formalism allows us to take effects due to the neutrino background
into account in a systematic and elegant way.

In section \ref{sec:FTQFT}, we evaluate the damping of a \UHEnu  travelling through the \CnuB in 
terms of the imaginary part of the neutrino self energy.
From the damping we determine the absorption probability for an \UHEnu emitted at a given 
redshift.

We then present in section \ref{sec:applications} some
illustrations of our calculation in realistic physical contexts.
First, the shape and position of the
absorption lines in the spectrum of \UHEnu from interactions with the \CnuB depend on the mass of 
the neutrinos and on the type and
distribution of sources for \UHEnu. We explore various combinations of
parameters to investigate the differences between the
finite temperature calculation and previous approximations and to determine the regimes in which 
those approximations break down and thermal effects become
significant. Then we further illustrate our results in the context
of relic neutrino clustering for different hypothesis on the
density and scale of the clusters. Conclusions are drawn in section~\ref{sec:conclusions}.

\section{Damping rate and transmission probability of an \UHEnu}
\label{sec:FTQFT}
\subsection{Self-energy in the relic neutrinos thermal background}

\begin{figure}[b!]
\begin{center}
\begin{picture}(170,130)(-85,-30)
\ArrowLine(40,0)(80,0) \Text(60,-10)[c]{$\nu(k)$}
\ArrowLine(-40,0)(40,0) \Text(15,-10)[c]{$\nu(p)$}
\ArrowLine(-80,0)(-40,0) \Text(-60,-10)[cr]{$\nu(k)$}
\PhotonArc(0,0)(40,0,180){4}{6.5} \GCirc(0,43){5}{0}
\Text(15,50)[cb]{$Z(q)$} \Line(0,-30)(0,80)
\end{picture}
\caption{\footnotesize{Feynman diagram for the one-loop self-energy of an UHE neutrino due to a 
Z-boson exchange with an (anti-)neutrino from the relic background; the blob on the Z propagator 
indicates that we use the dressed propagator and the cut is to select the imaginary part of the 
diagram.}} \label{fig:feynman}
\end{center}
\end{figure}
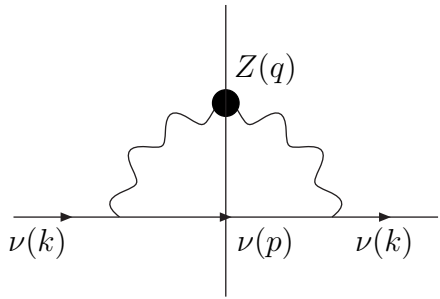

The dispersion relation of a particle that propagates through a
medium is determined from the linear part of the effective field
equation. In momentum space, for a neutrino with four-momentum
$k^\mu$ and mass $m_\nu$, it takes the form
\begin{equation}
\label{eq:EOMDirac} (\slache{k} - m_\nu - \Sigma)\  \psi =
0,
\end{equation}
where $\Sigma$ corresponds to the retarded self-energy and embodies the background effects.
For Dirac neutrinos, the chiral nature of neutrino interactions implies that, to one-loop order,
\begin{equation}
\Sigma=(a\, \slache{k} + b\, \slache{u})\ L,
\end{equation}
where $L=(1-\gamma_5)/2$ and $u^\mu$ is the velocity four-vector
of the medium; in its own rest frame $u^\mu = (1,\vec{0})$ and
$k^\mu = (\mathcal{E},\vec{K})$. The coefficients $a$ and $b$ are
complex functions of the scalars
\begin{equation}
\mathcal{E} = k.u, \hspace{2cm} K=\sqrt{\mathcal{E}^2 - k^2},
\end{equation}
with $K = \bigl|\vec{K}\bigr|$.
 In the present context $\Sigma$ corresponds to the Feynman diagram of fig.~\ref{fig:feynman}, 
where the loop contains a relic (anti)neutrino from the thermal bath, with four-momentum  
$p^\mu=(E_p,\vec{p})$, and a Z boson with a blob indicating that we consider its decay width to 
all possible channels.

A consequence of the presence of a self-energy term in the equation of motion is to modify the 
dispersion relation of the incoming neutrino to
\begin{equation}
\mathcal{E}_K = \mathcal{E} _r - i\frac{\gamma}{2},
\end{equation}
where $\mathcal{E}_K$, $\mathcal{E}_r$ and $\gamma$ are functions of $K$. The real part, 
$\mathcal{E}_r$,
is in general not equal to $\sqrt{K^2 + m^2}$ but has some additional correction reflecting the 
dispersive interactions that can take place in the medium, while the imaginary part corresponds to 
the damping factor, or else said, to the total reaction rate~\cite{jcdo95}. 
For a constant damping, the survival probability of an UHE neutrino travelling through the relic 
neutrino background is
\begin{equation}
\label{eq:Pt}
P_\mathrm{T}(\tau)=e^{-\gamma\,\tau}
\end{equation}
as a function of the propagation time $\tau$. 

The damping factor is directly related to the imaginary part of the self-energy $\Sigma_i$.
In the real-time formalism of the  FTFT, both the propagators and
the self-energies become $2\times 2$ matrices. As shown in~\cite{nieves95},
$\Sigma_i$ can be expressed in terms of the 
off-diagonal elements of the self-energy matrix as:
\begin{equation}
\label{eq:sigmai} \Sigma_i=\frac{i}{2}\left(\Sigma_{12} -
\Sigma_{21} \right).
\end{equation}

The vertices of the theory are doubled compared to the
vacuum case; the subscript 1 denotes the normal vertices of the
Standard Model while the vertices labelled 2 get an extra minus sign. 
In our case, the expressions for $\Sigma_{12}$ as calculated
from the Feynman diagram of fig.~\ref{fig:feynman} is

\begin{equation}
\label{eq:sigmapq}
-i\ \Sigma_{12}=(\frac{g}{2\cos\theta_\mathrm{W}})^2\ \int\frac{d^4p}{(2\pi)^3}\ 
i\,D_{21}^{\nu\mu}(q)\ \gamma_\mu\ L\ i\,S_{12}(p)\ \gamma_\nu\ L,
\end{equation}
with \(D^{\mu\nu}_{21}(q)\) and \(S_{12}(p)\) denoting the corresponding
element of the Z boson and neutrino propagator, respectively. One
obtains \(\Sigma_{21}\) by interchanging the indices 1 and 2 everywhere in
eq.~(\ref{eq:sigmapq}).

Since the temperature of the medium is small compared to the boson
mass, we may discard the thermal contribution to the Z propagator.
Consequently, we have 
\begin{eqnarray}
D_{12}^{\mu\nu}(q)&=& 2i\ \im D^{\mu\nu}(q)\ \theta(-q.u),
 \label{eq:D12}\\
D_{21}^{\mu\nu}(q)&=& 2i\ \im D^{\mu\nu}(q)\ \theta(q.u), \label{eq:D21}
\end{eqnarray}
where $\theta$ is the step function and $D^{\mu\nu}(q)$ is the vacuum
propagator for the Z boson. For $D^{\mu\nu}(q)$ we
adopt the usual prescription around the
resonance in the unitary norm~\cite{LEPZ,bardin88,borrelli90}:
\begin{equation}
\label{eq:Zpropagator} D^{\mu\nu}(q) = \frac{-g_{\mu\nu}+
\frac{q_\mu q_\nu}{M^2}}{q^2-M^2+iq^2\frac{\Gamma}{M}}.
\end{equation}
 Here $M$ is the Z mass and $\Gamma$ is the total width for Z decaying to
 fermion pairs. At lowest order \(\Gamma\) can  be expressed, neglecting fermion masses, as:
\begin{equation}
\label{eq:GammaZ}
\Gamma  = \frac{\sqrt{2} G_\mathrm{F}\ M^3}{6\pi} 
\sum_f N^\mathrm{c}_f \Bigg((I^3_f)^2-2I^3_fQ_f\sin^2\theta_\mathrm{W} +2 Q_f^2 
\sin^4\theta_\mathrm{W}\Bigg),
\end{equation}
where \(G_F\) is the Fermi constant, $N^\mathrm{c}_f=1(3)$ for leptons (quarks), and $I^3_f$ and 
$Q_f$ are the fermion isospin and charge, respectively~\cite{LEPZ}. The numerical values for 
$\Gamma$ and $M$ are taken from~\cite{pdb}. We assume that the Z boson decays in vacuum.

The propagators for the relic neutrino entering eq.~(\ref{eq:sigmapq}) are
 \begin{eqnarray}
S_{12}(p)&=& 2\pi i\ \delta(p^2-m_\nu^2)\ \left[\eta_\mathrm{F} - \theta(-p.u)\right]\ 
(\slache{p}+m_{\nu}), \\
S_{21}(p)&=& 2\pi i\ \delta(p^2-m_\nu^2)\ \left[\eta_\mathrm{F} -
\theta(p.u)\right]\ (\slache{p}+m_{\nu}),
\end{eqnarray}
where
\begin{eqnarray}
\label{eq:eta}
\eta_\mathrm{F} &=& \theta(p.u)\,f_\nu(P) \ + \ \theta(-p.u)\,f_\nubar(P)\\
 &=& \frac{\theta(p.u)}{e^{\beta P-\alpha}+1}\ +\ \frac{\theta(-p.u)}{e^{\beta P+\alpha}+1},
\end{eqnarray}
with \(P\) denoting the magnitude of the
three-momentum in the rest-frame of the background.
The distribution of background relic (anti)neutrinos \(f_\nu(P)\) (\(f_{\bar\nu}(P)\))
assumes a relativistic Fermi-Dirac form, with 
$\beta=1/T_\nu$ and $\alpha = \mu/T_{\nu}$, $\mu$ being the chemical potential for neutrinos and  
$T_{\nu}$ the temperature of the relic neutrino bath. It is worth remarking here that, although 
relic neutrinos are not relativistic anymore at present time, their distribution maintains a 
relativistic form since their
interactions froze out at their
decoupling time, corresponding to $T_{\nu d}\sim 1$ MeV~\cite{langacker83}.

After replacing the expressions above in eq.~(\ref{eq:sigmai}), expanding the delta function and 
neglecting contributions of order $m_\nu^2/M^2$ coming from the term proportional to $q_\mu q_\nu$ 
in (\ref{eq:Zpropagator}), one gets that

\begin{eqnarray}
\label{eq:Tmu}
\Sigma_i &=& \left(\frac{g}{2\cos\theta_\mathrm{W}}\right)^2\ \int \frac{d^3 P}{(2\pi)^3}\ 
\frac{\gamma_\mu \slache{p}\,\gamma_\nu}{2E_p}L\ \left\{\im \left[D^{\mu\nu}(p+k)\right]\, 
f_\nubar(P)\ +\right. \nonumber\\
&& \nonumber \\
&& \!\!\!\!\!\!\!\!\!\!\!\!\!\!\!\!\!\!\! + \ \left. \im 
\left[D^{\mu\nu}(p-k)\right]\,\left[\theta(\mathcal{E}-E_p)(1-f_\nu(P))\, +\, 
\theta(E_p-\mathcal{E})\, f_\nu(P)\right]\right\},
\end{eqnarray}
where the integral is evaluated in the rest frame of the medium and we
replaced $p^0 $ by $E_p=(P^2 + m_\nu^2)^{1/2}$ everywhere. 

The first term corresponds to the resonant production of a Z boson through $\nu-\nubar$ 
annihilation; the factor of $f_\nubar(P)$ reflects the Pauli blocking acting on the antineutrinos 
in the background. The second and third terms correspond to the emission of a Z-boson respectively  
by the incoming neutrino ($\mathcal{E}_r > E_p$) or by the background antineutrino ($E_p > 
\mathcal{E}_r$). Both processes are kinematically forbidden and we drop them from this point on.

\subsection{Ultrarelativistic approximation}

Let us  assume that the neutrino travelling through the relic neutrino thermal bath is 
ultrarelativistic and  that we can neglect the background effects
on its energy, {\it i.e.}, $\mathcal{E}_r \simeq K$. In that case,
we can use the following expression for the damping:
\begin{equation}
\label{eq:gammaUR} \gamma \simeq -2\  \im b(K,K)  =
-\frac{1}{K}\left.\mathop{\mathrm{Tr}}(\slache{k}\Sigma_i)\right|_{\mathcal{E}_r=K}.
\end{equation}
This expression was derived in~\cite{jcdo95} for the case of
a massless fermion, but it can be shown 
that it remains valid for a massive one in the relativistic limit.
The term proportional to $q_\alpha q_\beta$ in the Z propagator gives contributions of order 
$m_\nu^2/M^2$; if we discard them, then the damping
rate corresponding to the $\nu\nubar$ annihilation process reads
\begin{eqnarray}
\nonumber
\gamma (K)&=& \frac{g^2}{\cos^2\theta_\mathrm{W}} \frac{\Gamma}{M} \int
\frac{d^3 P}{(2 \pi)^3}\frac{f_\nubar(P)}{2 K E_p}\
\frac{(k+p)^2 (k.p)}{(1+\xi)(k+p)^4 - 2 M^2 (k+p)^2 +
M^4},
\end{eqnarray}
where $\xi = \Gamma^2/M^2 \ll 1$ and $p^0 = E_p$.  The
previous formula can be conveniently rewritten in terms of the
momentum
integration of the cross-section for the process $\nu \nubar \to Z$, weighted by the corresponding 
statistical factor:  
\begin{equation}
\label{eq:gammasigma} \gamma(K) = \int_0^\infty \frac{\d P}{2\pi^2}
\ P^2 \ f_\nubar (P) \ \sigma_{\nu\nubar} (P,K),
\end{equation}
where 
\begin{equation}
\label{eq:sigma} \sigma_{\nu\nubar}(P,K)=
\frac{G_\mathrm{F}}{\sqrt{2}}\frac{\Gamma M}{2 K^2} \frac{1}{P
E_P}\int_{s_-}^{s_+} \d s \
\frac{s(s-2m_\nu^2)}{\left(s-M^2\right)^2+\xi s^2}.
\end{equation}
Neglecting the chemical
potential, the distribution function of the relic anti-neutrinos is:
\begin{equation}
\label{eq:FDdistri}
f_\nubar(P)=f_\nu(P)=\frac{1}{e^{P/T_{\nu}}+1},
\end{equation}

The integration variable $s
= (k+p)^2$ corresponds to the total energy in the
center-of-mass and
\begin{equation}
s_\pm = 2m_\nu^2 + 2K(E_p\pm P). \label{eq:spm}
\end{equation}

The integral  in eq.~(\ref{eq:sigma}) can be done in a closed form.
For $m_\nu \ll M,K$, we get
\begin{eqnarray}
\sigma_{\nu\nubar}(P,K) &=& \frac{2\sqrt{2}G_\mathrm{F}\Gamma M}{2KE_p}
\left\{ \frac{1}{1+\xi} + \frac{M^2}{4KP(1+\xi)^2}\right. \nonumber \\
&& {}\times \ln\left(\frac{(1+\xi)4K^2(E_p + P)^2 -
4M^2K(E_p+P)+M^4}{(1+\xi)4K^2(E_p - P)^2 -
4M^2K(E_p-P)+M^4}\right) \nonumber \\
&& {}+ \frac{1-\xi}{(1+\xi)^2}\frac{M^3}{4KP 
\Gamma}\left[\arctan\left(\frac{2K(1+\xi)(E_p+P)-M^2}{\Gamma
M}\right)
\right. \nonumber \\
&&\left.\left.
\hspace{2.5cm}-\arctan\left(\frac{2K(1+\xi)(E_p-P)-M^2}{\Gamma
M}\right)\right]\right\}. \label{eq:sigmanubar}
\end{eqnarray}

 \begin{figure}[ht!]
\begin{center}
 \psfrag{27}[c]{\tiny \phantom{n} \raisebox{0.1cm}{$10^{27}$}}
 \psfrag{26}[c]{\tiny \phantom{n} \raisebox{0.1cm}{$10^{26}$}}
 \psfrag{25}[c]{\tiny \phantom{n} \raisebox{0.1cm}{$10^{25}$}}
 \psfrag{24}[c]{\tiny \phantom{n} \raisebox{0.1cm}{$10^{24}$}}
 \psfrag{23}[c]{\tiny \phantom{n} \raisebox{0.1cm}{$10^{23}$}}
 \psfrag{22}[c]{\tiny \phantom{n} \raisebox{0.1cm}{$10^{22}$}}
 \psfrag{21}[c]{\tiny \phantom{n} \raisebox{0.1cm}{$10^{21}$}}
 \psfrag{20}[c]{\tiny \phantom{n} \raisebox{0.1cm}{$10^{20}$}}
 \psfrag{19}[c]{\tiny \phantom{n} \raisebox{0.1cm}{$10^{19}$}}
 \psfrag{-36}[c]{\tiny \phantom{i} \raisebox{0.1cm}{$10^{^{-36}}$}}
 \psfrag{-35}[c]{\tiny \phantom{i} \raisebox{0.1cm}{$10^{^{-35}}$}}
 \psfrag{-34}[c]{\tiny \phantom{i} \raisebox{0.1cm}{$10^{-34}$}}
 \psfrag{-33}[c]{\tiny \phantom{i} \raisebox{0.1cm}{$10^{-33}$}}
 \psfrag{-32}[c]{\tiny \phantom{i} \raisebox{0.1cm}{$10^{-32}$}}
 \psfrag{-31}[c]{\tiny \phantom{i} \raisebox{0.1cm}{$10^{-31}$}}
 \psfrag{X}[c]{\tiny \raisebox{-0.5cm}{$K\ [\mathrm{eV}]$}}
 \psfrag{Y}[c]{\tiny \raisebox{-1cm}{$\sigma_{\nu\nubar}\ [\mathrm{cm}^2]$}}
 \psfig{figure=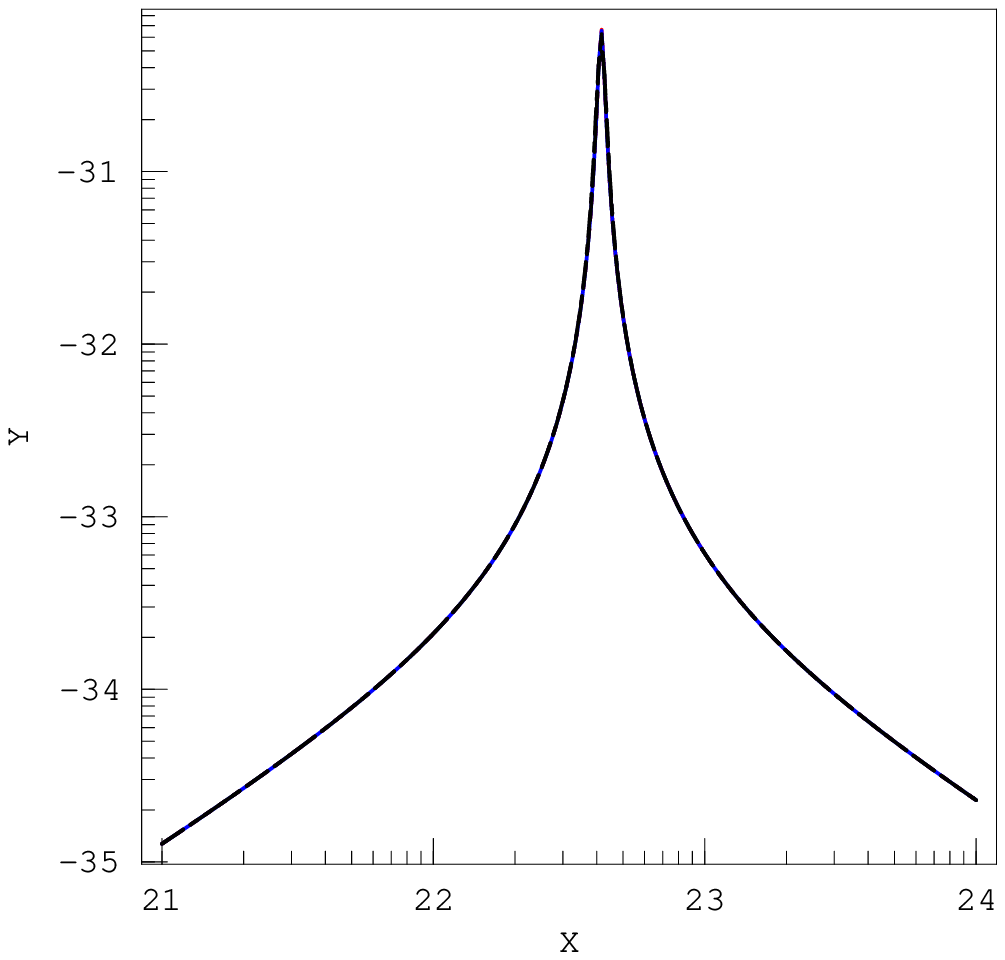,height=4cm,width=5.5cm,angle=0}
 \hfill
 \psfig{figure=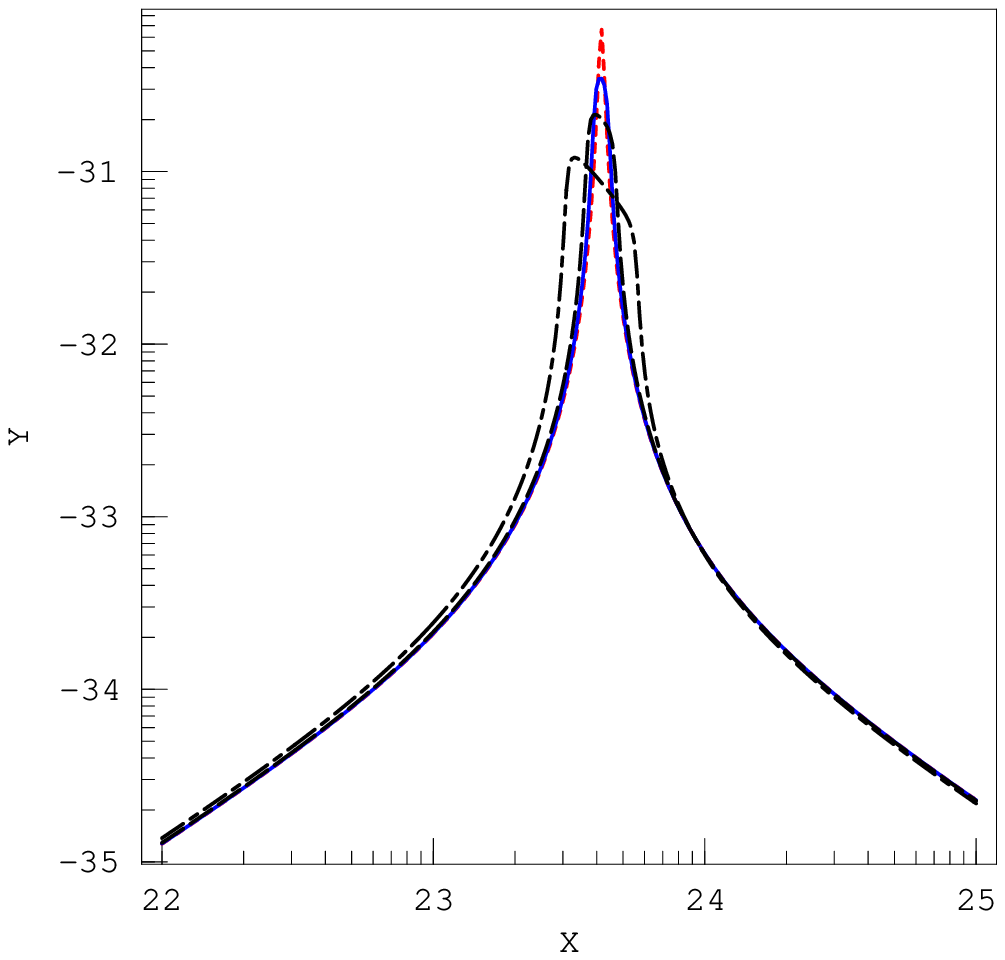,height=4cm,width=5.5cm,angle=0}
 \hfill \\[3ex]
 \psfig{figure=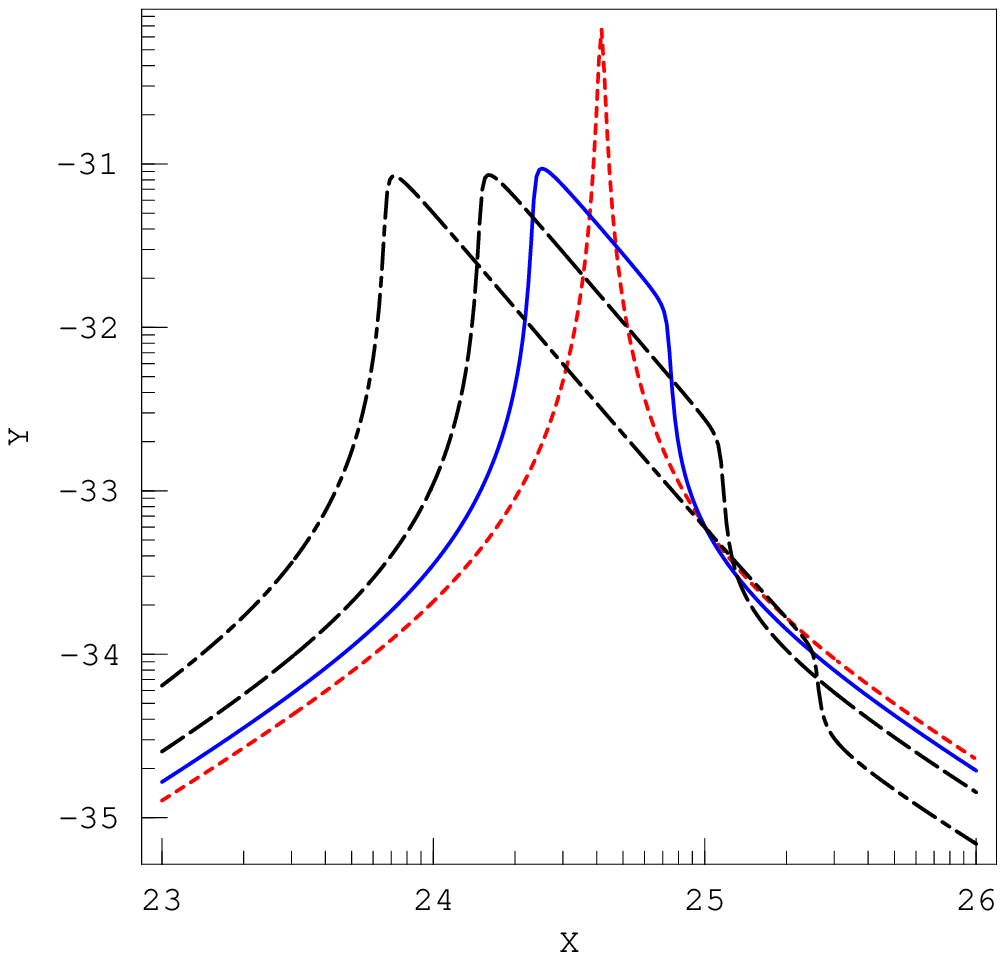,height=4cm,width=5.5cm,angle=0}
 \hfill
 \psfig{figure=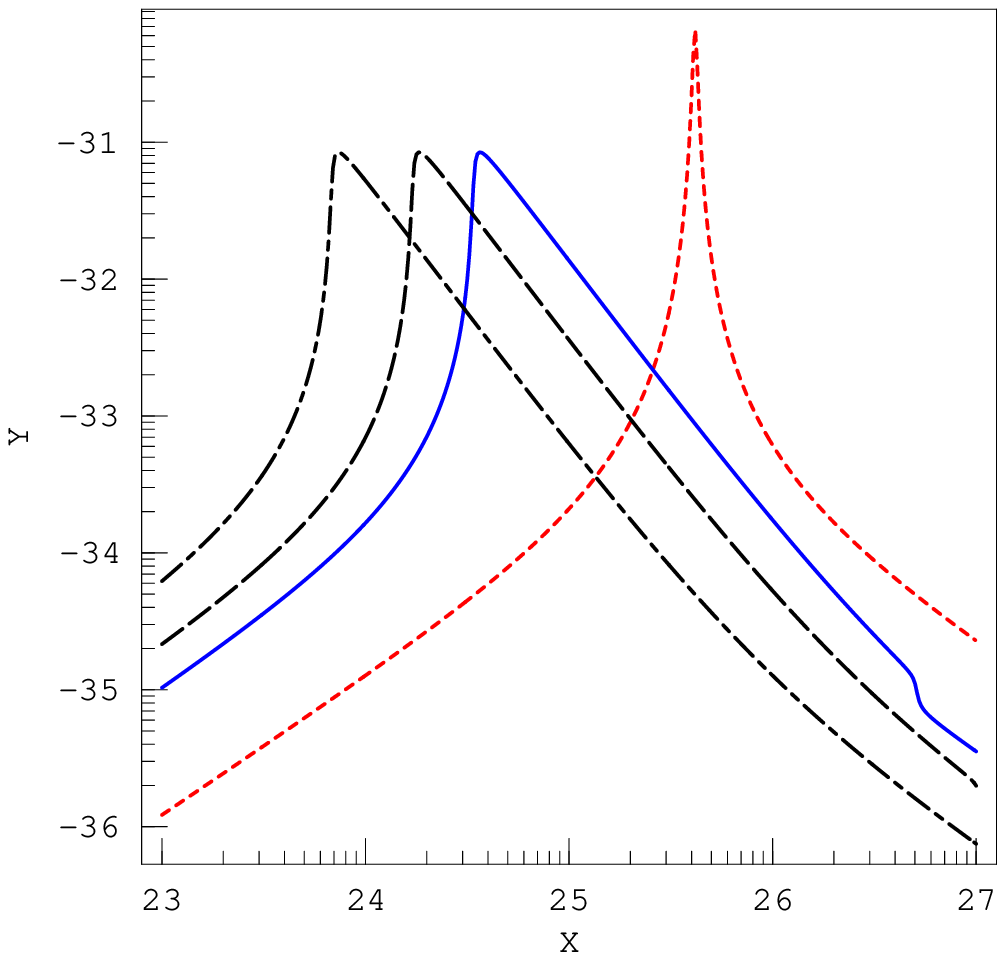,height=4cm,width=5.5cm,angle=0}
 \hfill
 \caption{\footnotesize{Cross-section $\sigma_{\nu\nubar}(P,K)$, in $cm^2$, as given by 
eq.~(\ref{eq:sigmanubar}), as a function of the energy of the incident neutrino, $K$, and for 
neutrino masses of $10^{-1}$, $10^{-2}$, $10^{-3}$ and $10^{-4}$ eV (from left to right and top to 
bottom). Each plot shows four curves corresponding to different relic neutrino momenta; the dotted 
(red) curve corresponds to the approximation of relic neutrinos at rest, $P=0$; the continuous 
(blue) curve is for $P_\mathrm{rms}=\sqrt{\left<P^2\right>}\simeq 6.08\times 10^{-4}$ eV and the 
dashed and dot-dashed (black) curves are respectively for  $2 P_\mathrm{rms}$ and
$5 P_\mathrm{rms}$.}} \label{fig:sigma}
 \end{center}
 \end{figure}

In fig.~\ref{fig:sigma}, we plot the cross-section as a function of the \UHEnu energy $K$, for 
neutrino masses $m_\nu$ ranging from $10^{-1}$ eV to $10^{-4}$ eV. Each curve corresponds to a 
particular value of the relic neutrino momentum $P$. As expected, the thermal effects become more 
and more important as the ratio between the neutrino mass and the \CnuB temperature (which we take 
here as $T_{\nu 0} \simeq 1.69 \times 10^{-4}$ eV) decreases. We also compare with the value of 
the cross-section in the approximation of relic neutrinos at rest, $\sigma_{\nu\nubar}(0,K)$, 
which is obtained by taking the limit of $\sigma_{\nu\nubar}(P,K)$ for $P\to 0$. As long as the 
neutrino mass is sufficiently
large, that is $m_\nu > 0.01$ eV, the cross-section does not vary
significantly over the relevant range of $P$, {\it i.e.}, the range of momenta selected by the 
\CnuB distribution function around the maximum of $P^2 f_\nu(P)$ at $P_\mathrm{max} \simeq 3.75 
\times
10^{-4}$ eV. In this case, the position of the peak corresponds to the resonant energy 
$K_\mathrm{res}^0=M^2/2m_\nu$ for
target neutrinos at rest. For smaller masses, the peak in the cross-section gets broader and 
shifts to smaller \UHEnu energies for increasing relic neutrino momentum. The maximum value is 
approximately constant over the relevant range of momenta, and is about one order of magnitude 
lower than $\sigma_{\nu\nubar}(0,K_\mathrm{res})$.

A simpler expression for $\sigma_{\nu\nubar}$ can be obtained using the mean value theorem by 
evaluating the integrand in eq.~(\ref{eq:sigma}) at the midpoint of the integration interval, 
$\bar{s} \simeq 2KE_p$. Taking into account that
$s_+ - s_- = 4 K P$ we get the following expression:

\begin{equation}
\label{eq:barsigmabar}
\bar{\sigma}_{\nu\nubar} (\bar{s})=2\sqrt{2} G_\mathrm{F} \Gamma M \ \   
\frac{\bar{s}}{\left(\bar{s} - M^2\right)^2 + \xi \bar{s}^2},
\end{equation}
where all the dependence on K and P is implicit in $\bar{s}=2 K
\sqrt{m^2+P^2}$.

We can use these expressions to evaluate the integral in
eq.~(\ref{eq:gammasigma}) numerically. 
A common approximation consists in neglecting the thermal motion of the background neutrinos (see 
for example~\cite{roulet93,ringwaldnew}). The corresponding expression can be recovered in our 
formalism by evaluating the cross-section (\ref{eq:barsigmabar}) in $P=0$, {\it i.e.}, in $\bar 
s_0 = 2m_\nu K$, or equivalently by taking the limit for $P\to 0$ of the general expression 
(\ref{eq:sigmanubar}). The damping reads in this case:
\begin{eqnarray}
\gamma^0(K) &=& \bar{\sigma}_{\nu\nubar}(\bar s_0)\ n_\nu \\
 &=& 2\sqrt{2} G_\mathrm{F} \Gamma M\ \frac{2 K m}{4(1+\xi)K^2 m^2 - 4M^2K m +M^4} \ n_\nu, 
\label{eq:gammaRoulet}
\end{eqnarray}
with
\begin{equation}
  \label{eq:nutotal}
  n_\nu = n_\nubar = \int_0^\infty \frac{\d P}{2\pi^2} P^2 f_\nu(P).
\end{equation}

Substituting the full expression for the total Z width to fermions, eq.~(\ref{eq:GammaZ}) into 
eq.~(\ref{eq:gammaRoulet}), one can easily recover the result
of~\cite{roulet93}. The expressions of~\cite{ringwaldnew} can be obtained by further evaluating 
the cross-section at the pole of the resonance $2 m K_\mathrm{res} = M^2$ (narrow-width 
approximation).

\begin{figure}[ht!]
   \begin{center}
 \psfrag{0}[c]{\tiny  \phantom{i}\raisebox{0.1cm}{$0$}}
 \psfrag{10}[c]{\tiny \phantom{ni} \raisebox{0.1cm}{$1$}}
 \psfrag{20}[c]{\tiny \phantom{ni} \raisebox{0.1cm}{$2$}}
 \psfrag{30}[c]{\tiny \phantom{ni} \raisebox{0.1cm}{$3$}}
 \psfrag{40}[c]{\tiny \phantom{ni} \raisebox{0.1cm}{$4$}}
 \psfrag{50}[c]{\tiny \phantom{ni} \raisebox{0.1cm}{$5$}}
 \psfrag{1}[c]{\tiny \phantom{n}\raisebox{0.1cm}{$1$}}
 \psfrag{2}[c]{\tiny \phantom{i}\raisebox{0.1cm}{$2$}}
 \psfrag{3}[c]{\tiny \phantom{i}\raisebox{0.1cm}{$3$}}
 \psfrag{4}[c]{\tiny \phantom{i}\raisebox{0.1cm}{$4$}}
 \psfrag{5}[c]{\tiny \phantom{i}\raisebox{0.1cm}{$5$}}
 \psfrag{6}[c]{\tiny \phantom{i}\raisebox{0.1cm}{$6$}}
 \psfrag{7}[c]{\tiny \phantom{i}\raisebox{0.1cm}{$7$}}
 \psfrag{8}[c]{\tiny \phantom{nnnni}\raisebox{0.15cm}{$8 \ 10^{22}$}}
 \psfrag{77}[c]{\tiny \phantom{nnnnni}\raisebox{0.15cm}{$7 \ 10^{25}$}}
 \psfrag{88}[c]{\tiny \phantom{nnnnni}\raisebox{0.15cm}{$8 \ 10^{23}$}}
 \psfrag{888}[c]{\tiny \phantom{nnnnnni}\raisebox{0.15cm}{$8 \ 10^{24}$}}
 \psfrag{X}[c]{\tiny \raisebox{-0.5cm}{$K_0 [\mathrm{eV}]$}}
 \psfrag{Y}[c]{\tiny {$\gamma_{\nu\nubar} [10^{-34}\ \mathrm{eV}]$}}

 \psfig{figure=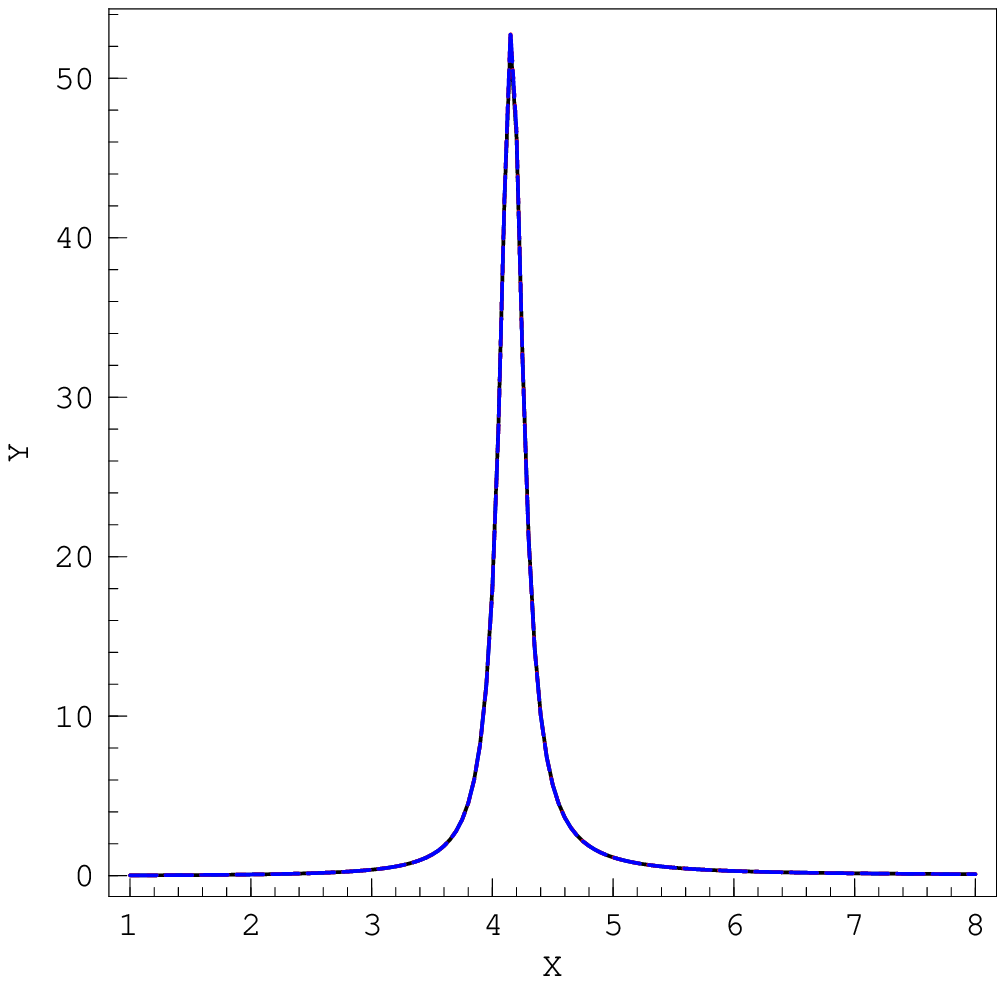,height=4cm,width=5cm,angle=0}
 \hspace{2cm}
 \psfig{figure=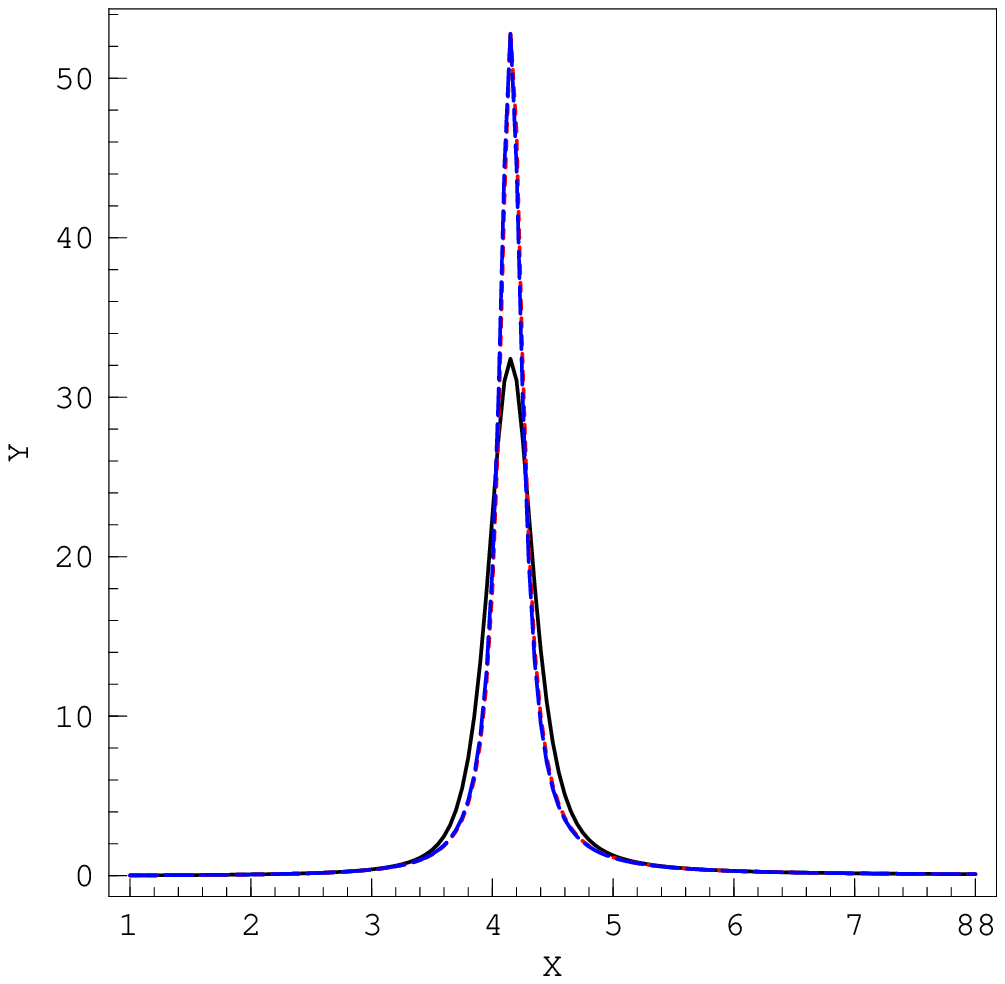,height=4cm,width=5cm,angle=0} \\[3ex]
 \psfig{figure=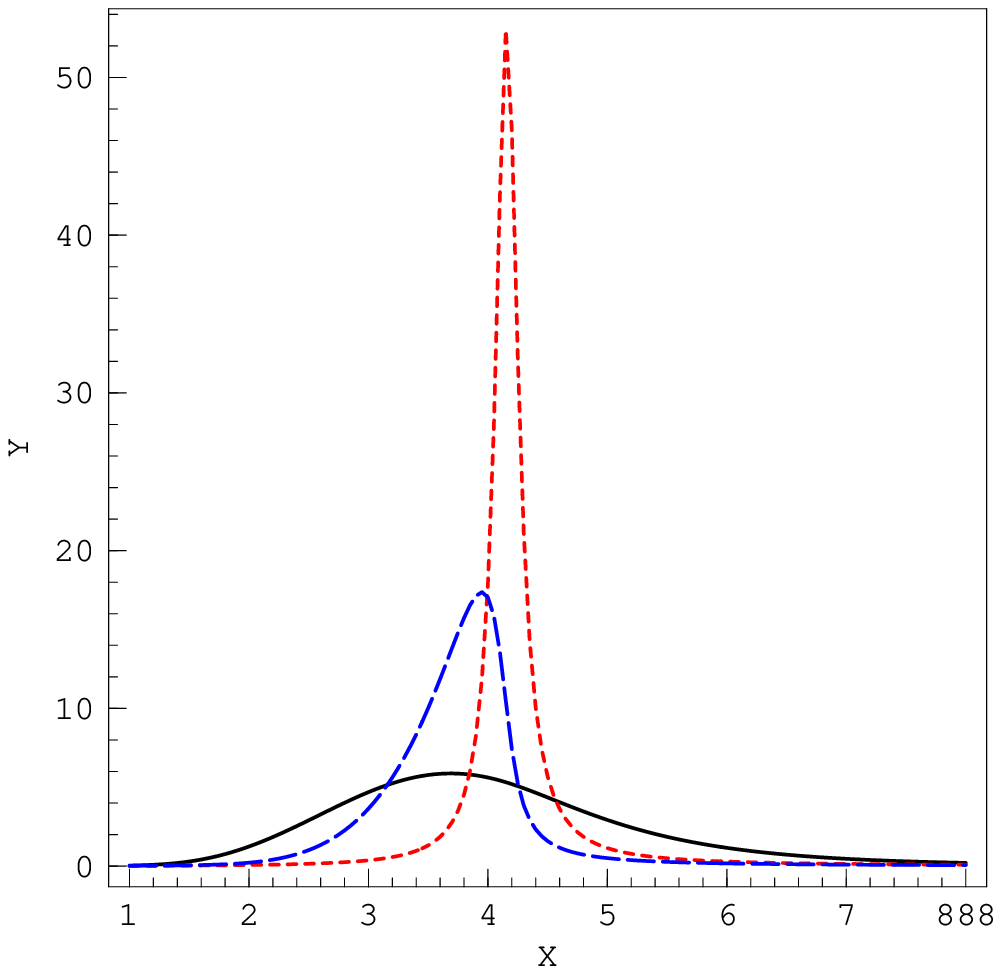,height=4cm,width=5cm,angle=0}
 \hspace{2cm}
 \psfig{figure=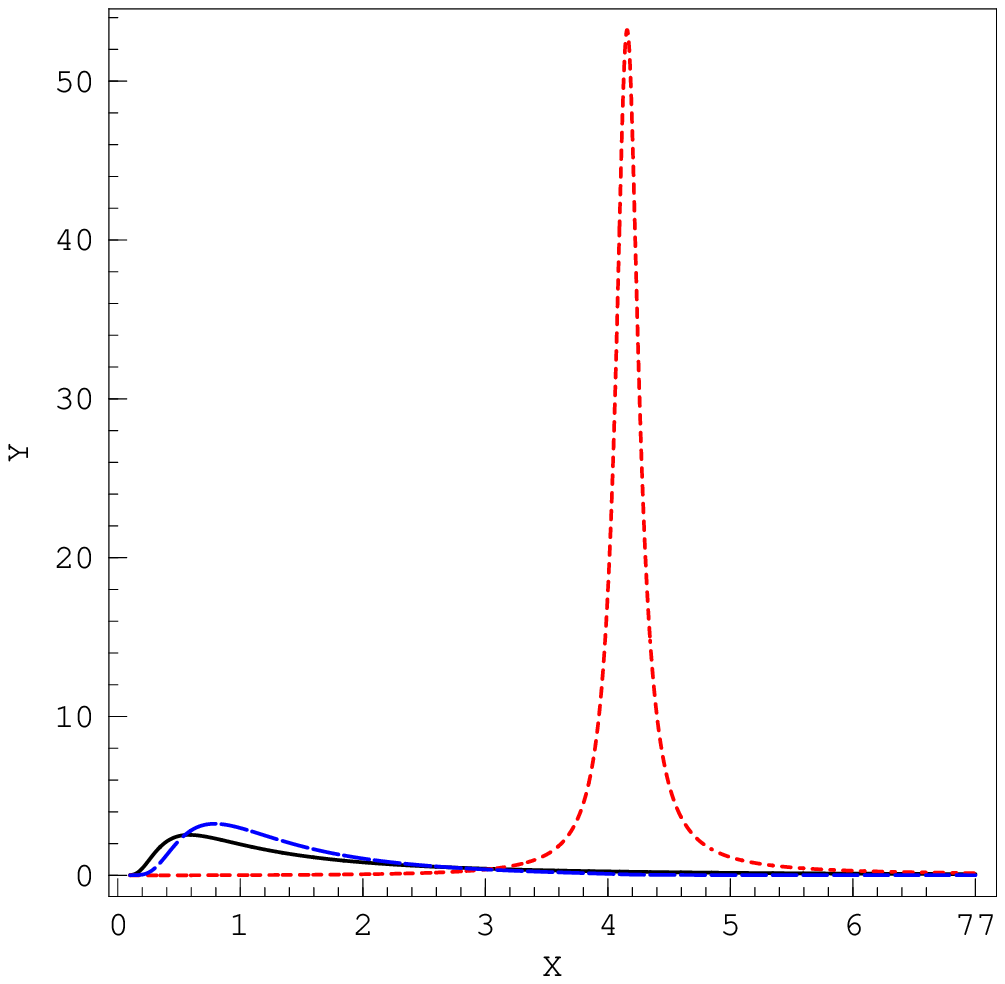,height=4cm,width=5
cm,angle=0}
 \caption{\footnotesize{Damping factor due to $\nu - \nubar$ annihilation, in eV, as a function of 
UHE neutrino energy $K$; each plot corresponds to a given value of the neutrino mass ($10^{-1}$ 
eV, $10^{-2}$ eV, $10^{-3}$ eV, $10^{-4}$ eV from left to right and from top to bottom) and 
displays the integrated damping, $\gamma_{\nu\nubar}$, from the full cross-section 
eq.~(\ref{eq:sigmanubar}) (black, continued curve) and from the approximated cross-section 
eq.~(\ref{eq:barsigmabar}) (blue, dashed curve), as well as the approximation for neutrinos at 
rest, $\gamma^0_{\nu\nubar}$, from  eq.~(\ref{eq:gammaRoulet}) (red, dotted curve). The three 
curves are superposed for $m_\nu = 0.1$ eV, and so are the two approximations for $m_\nu = 0.01$ 
eV.}}
\label{fig:damping}
 \end{center}
 \end{figure}

The corresponding results are plotted in fig.~\ref{fig:damping}, where we directly compare the 
approximations for the damping factor, eq.~(\ref{eq:gammaRoulet}) and
eq.~(\ref{eq:barsigmabar}), with the exact expression obtained from eqs.~(\ref{eq:gammasigma}) and 
(\ref{eq:sigmanubar}).
The comparison is done for different values of the neutrino mass. 
One notices that  the net effect of thermal
 broadening is a
 reduction of the damping, which affects the transmission probability and the depth and  shape of 
the absorption dips.

Two effects combine: the modification of the cross-section peak due to its dependence on $E_p$, 
and the presence of the thermal distribution which selects a range of relic neutrino momenta 
comparable to the temperature of the \CnuB. Both effects contribute to smearing out the resonance 
peak and shifting its maximum value from $K_\mathrm{res}$ to lower energies. As long as the ratio 
$m_\nu / T_\nu$ is $\gtrsim 100$, the position of the peak is not significantly affected and the 
net effect is to reduce the damping. For smaller masses, the effect of thermal broadening is much 
stronger and the damping gets spread over a larger range of \UHEnu energies, resulting in a worse 
definition of the absorption dip.

The comparison of the exact and approximate results for the damping in
fig.~\ref{fig:damping} illustrates that 
none of the approximations is valid over the range of neutrino masses
we consider. 
In the subsequent analysis we therefore keep on working with the full expression for the 
cross-section $\sigma_{\nu\nubar}(K,P)$, eq.~(\ref{eq:sigmanubar}).

\subsection{Transmission Probability across the relic neutrino
background}

High energy neutrinos can travel cosmological distances almost without interacting. To calculate 
the damping by the \CnuB one has to take the expansion of the Universe into account.
The standard procedure is to calculate the transmission probability by integrating the damping 
factor over the \UHEnu path, in terms of the redshift $z$, back to the source position 
$z_\mathrm{s}$. The formula (\ref{eq:Pt}) is then generalised as follows:
\begin{equation}
\label{eq:PTredshift}
P_\mathrm{T}(K_0,z_\mathrm{s})=\exp\left[{-\int_0^{z_\mathrm{s}} \frac{\d z}{H(z)(1+z)} 
\gamma_{\nu\nubar}(K_0(1+z))}\right],
\end{equation}
with the neutrino temperature \(T_\nu = T_{\nu0}(1+z)\). Both \(K_0\) and
\(T_{\nu0}\) refer to the quantities in today's Universe. For the Hubble factor we take 
$H(z)=H_0\sqrt{0.3(1+z)^3+0.7}$ as suggested by recent observations~\cite{spergel03}, with the 
numerical value of $H_0$ from~\cite{pdb}.

Eq.~(\ref{eq:PTredshift}) encompasses two effects due to the expansion of the Universe. First, the 
\UHEnu energy gets shifted, $K\to K_0 (1+z)$. This broadens the absorption dip even in the 
approximation of relic neutrinos at rest, since the resonance energy changes along the \UHEnu 
path. Second, the temperature of the \CnuB gets shifted, $T_{\nu}\to T_{\nu 0}(1+z)$, 
\texttt{i.e.}, the thermal bath of relic neutrinos is hotter at earlier times. This directly 
affects the ratio between $m_\nu$ and $T_\nu$ and results in a  modification of the absorption 
properties of the
\CnuB with respect to the \UHEnu{s} that cross it.

 \begin{figure}
\begin{center}
 \psfrag{27}[c]{\tiny \phantom{n} \raisebox{0.1cm}{$10^{27}$}}
 \psfrag{26}[c]{\tiny \phantom{n} \raisebox{0.1cm}{$10^{26}$}}
 \psfrag{25}[c]{\tiny \phantom{n} \raisebox{0.1cm}{$10^{25}$}}
 \psfrag{24}[c]{\tiny \phantom{n} \raisebox{0.1cm}{$10^{24}$}}
 \psfrag{23}[c]{\tiny \phantom{n} \raisebox{0.1cm}{$10^{23}$}}
 \psfrag{22}[c]{\tiny \phantom{n} \raisebox{0.1cm}{$10^{22}$}}
 \psfrag{21}[c]{\tiny \phantom{n} \raisebox{0.1cm}{$10^{21}$}}
 \psfrag{20}[c]{\tiny \phantom{n} \raisebox{0.1cm}{$10^{20}$}}
 \psfrag{1.}[c]{\tiny \phantom{ni} \raisebox{0.1cm}{$1$}}
 \psfrag{0.8}[c]{\tiny \phantom{ni} \raisebox{0.1cm}{$0.8$}}
 \psfrag{0.6}[c]{\tiny \phantom{ni} \raisebox{0.1cm}{$0.6$}}
 \psfrag{0.4}[c]{\tiny \phantom{ni} \raisebox{0.1cm}{$0.4$}}
 \psfrag{0.2}[c]{\tiny \phantom{ni} \raisebox{0.1cm}{$0.2$}}
 \psfrag{0.}[c]{\tiny \phantom{ni} \raisebox{0.1cm}{$0$}}
 \psfrag{X}[c]{\tiny \raisebox{-0.5cm}{$K_0 [\mathrm{eV}]$}}
 \psfrag{Y}[c]{\tiny {$P_\mathrm{T}$}}
 \psfig{figure=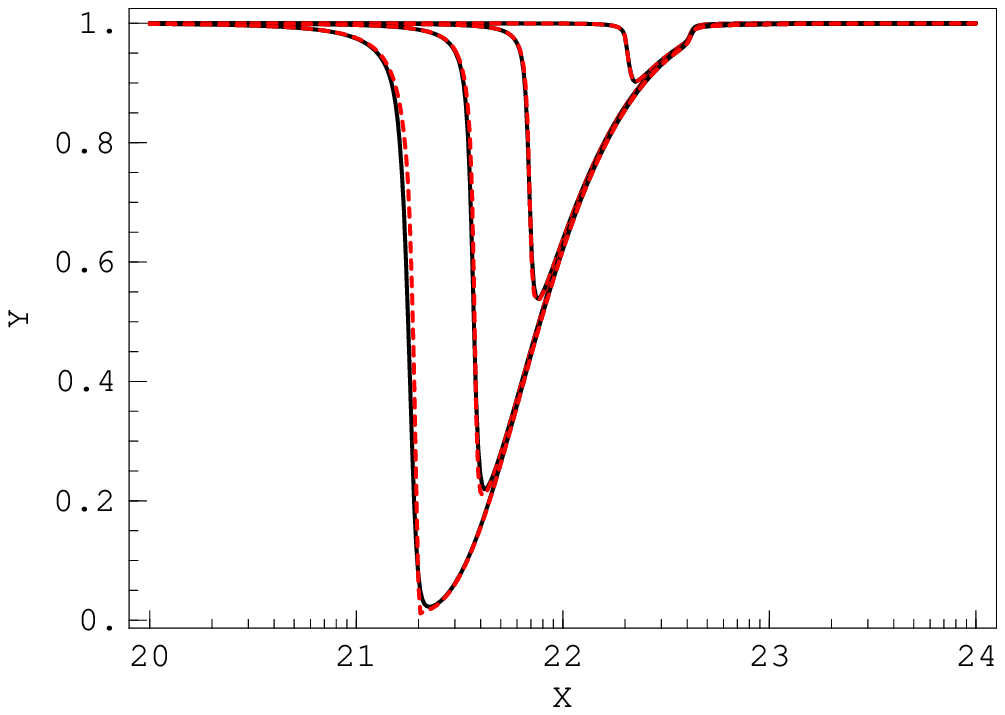,height=4cm,width=5cm,angle=0}
 \hfill
 \psfig{figure=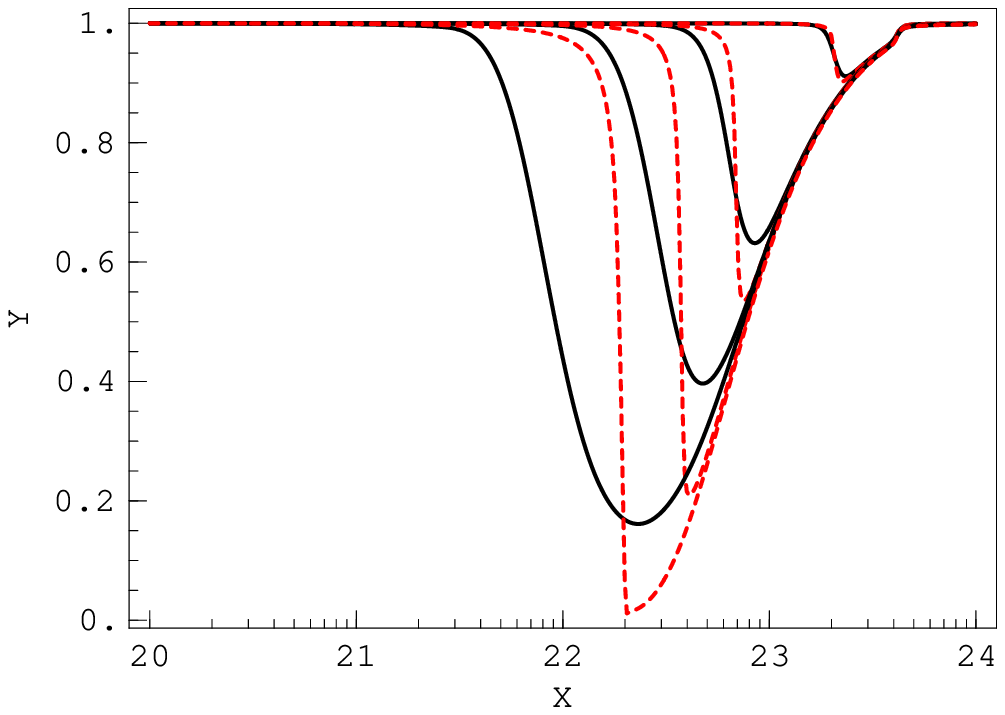,height=4cm,width=5cm,angle=0} \\[3ex]
 \psfig{figure=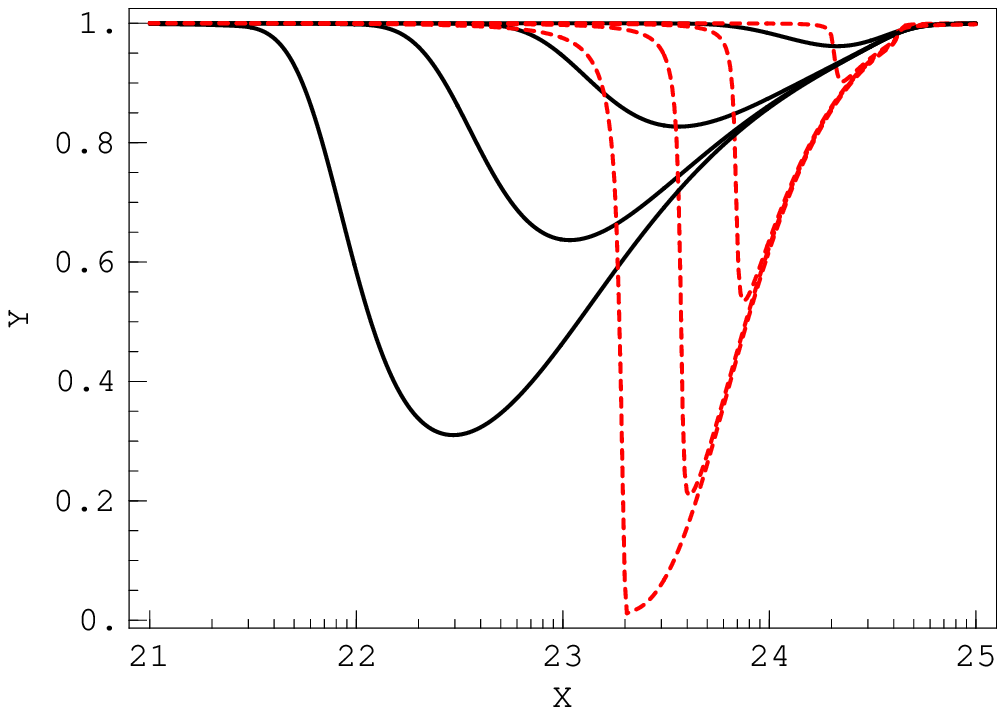,height=4cm,width=5cm,angle=0}
 \hfill
 \psfig{figure=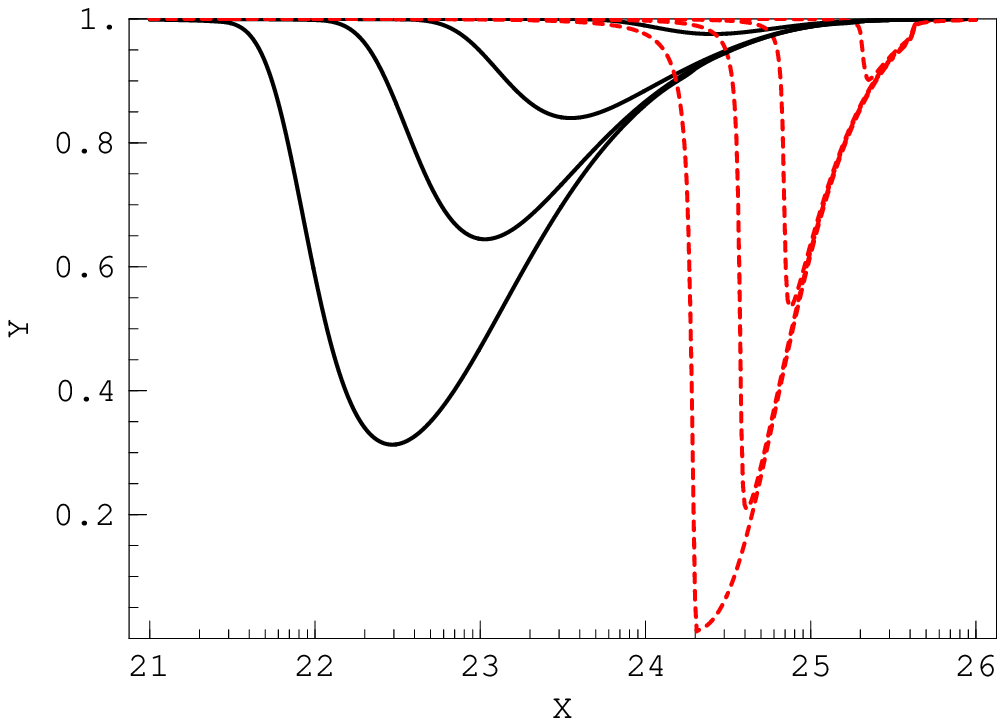,height=4cm,width=5
cm,angle=0}
 \caption{\footnotesize{Transmission probability $P_\mathrm{T}(K_0,z_\mathrm{s})$ as a function of 
the incident neutrino energy as detected on Earth, $K_0$, for an UHE neutrino source located at 
redshifts $z_\mathrm{s}=1$, $5$, $10$, $20$ (from top to bottom in each plot) and for a neutrino 
mass $m_\nu = 10^{-1}$, $10^{-2}$, $10^{-3}$, $10^{-4}$ eV. The continued, black curves 
corresponds to the full $\gamma_{\nu\nubar}$ from eqs.~(\ref{eq:gammasigma}) and 
(\ref{eq:sigmanubar}), while the dotted (red) curves are for the approximation of relic neutrinos 
at rest, $\gamma^0_{\nu\nubar}$, from eq.~(\ref{eq:gammaRoulet}). }} \label{fig:transmission}
 \end{center}
 \end{figure}
In fig.~\ref{fig:transmission} we show the transmission probability for an \UHEnu emitted at a 
fixed redshift \(z_\mathrm{s} = 1\), 5, 10 or 20, as a function of its present energy, $K_0$, and 
compare it with the results obtained in the
approximation of relic neutrinos at rest.

As long as $m_\nu/T_{\nu} \gtrsim 10^2$, the shape of the absorption dip is not affected by 
thermal broadening and is rather sharply delimited, at high energies, by the bare resonant energy 
for the propagating neutrino, $K_\mathrm{res} = M^2/(2m_\nu)$, and at low energies by the 
redshifted resonant energy $K_\mathrm{res}/(1+z)$. Evaluating the position of these points would 
in principle allow us to determine the value of the neutrino mass as well as the redshift at which 
the \UHEnu was emitted. As $m_\nu/T_{\nu}$ decreases, however, the absorption dips gets shallower 
and broader, which complicates the extraction of $m_\nu$ and $z_\mathrm{s}$. The position of the 
minimum transmission probability is also shifted to lower energies. The effect of thermal motion 
increases with the redshift $z$ since UHE neutrinos from more distant sources are emitted in 
hotter backgrounds.

 \section{Applications}
\label{sec:applications}
 The results presented so far deal with a monoenergetic source of UHE neutrinos located at a given 
redshift; let us now illustrate the potential effects of thermal motion on the process of 
absorption of UHE neutrinos in two more realistic contexts of physical relevance.
\subsection{Absorption lines in a realistic UHE neutrino flux}
\label{sec:fluxes}

\begin{figure}[htbp!]
\begin{center}

 \psfrag{26}[c]{\tiny \phantom{n} \raisebox{0.1cm}{$10^{26}$}}
 \psfrag{25}[c]{\tiny \phantom{n} \raisebox{0.1cm}{$10^{25}$}}
 \psfrag{24}[c]{\tiny \phantom{n} \raisebox{0.1cm}{$10^{24}$}}
 \psfrag{23}[c]{\tiny \phantom{n} \raisebox{0.1cm}{$10^{23}$}}
 \psfrag{22}[c]{\tiny \phantom{n} \raisebox{0.1cm}{$10^{22}$}}
 \psfrag{21}[c]{\tiny \phantom{n} \raisebox{0.1cm}{$10^{21}$}}
 \psfrag{20}[c]{\tiny \phantom{n} \raisebox{0.1cm}{$10^{20}$}}
 \psfrag{19}[c]{\tiny \phantom{n} \raisebox{0.1cm}{$10^{19}$}}
 \psfrag{1.}[c]{\tiny \phantom{ni} \raisebox{0.1cm}{$1$}}
 \psfrag{0.8}[c]{\tiny \phantom{ni} \raisebox{0.1cm}{$0.8$}}
 \psfrag{0.6}[c]{\tiny \phantom{ni} \raisebox{0.1cm}{$0.6$}}
 \psfrag{0.4}[c]{\tiny \phantom{ni} \raisebox{0.1cm}{$0.4$}}
 \psfrag{0.2}[c]{\tiny \phantom{ni} \raisebox{0.1cm}{$0.2$}}
 \psfrag{0.}[c]{\tiny \phantom{ni} \raisebox{0.1cm}{$0$}}
 \psfrag{X}[c]{\tiny \raisebox{-0.5cm}{$K_0 [\mathrm{eV}]$}}
 \psfrag{Y}[c]{\tiny {${\mathcal{F}_\nu}/{\mathcal{F}_{\nu 0}}$}}

 \psfig{figure=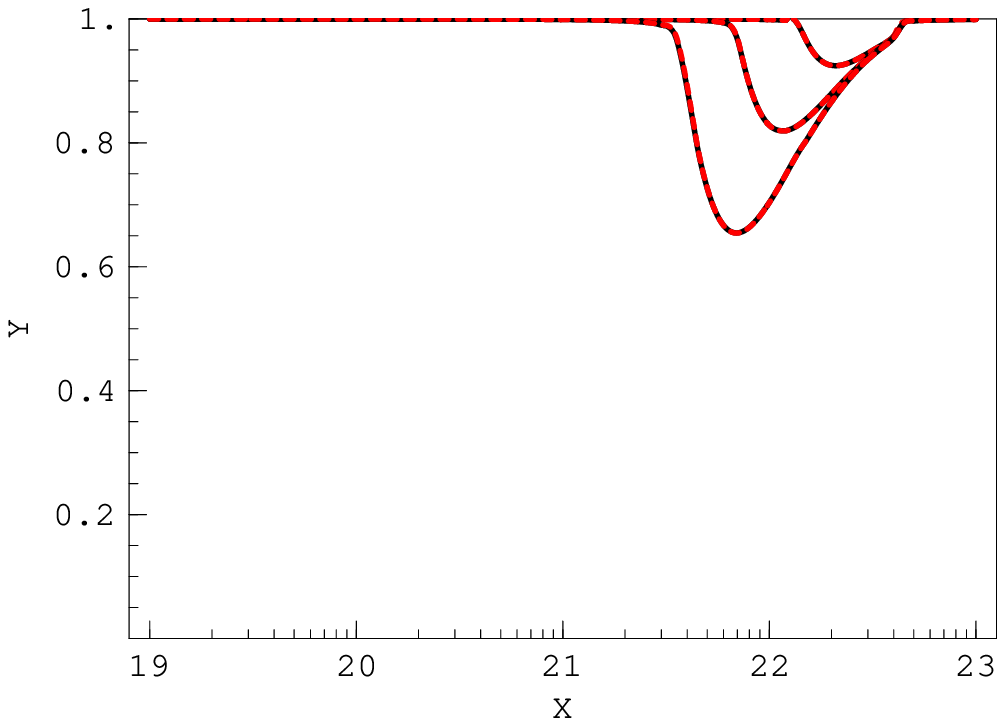,height=4.5cm,width=6cm,angle=0}
 \hfill
 \psfig{figure=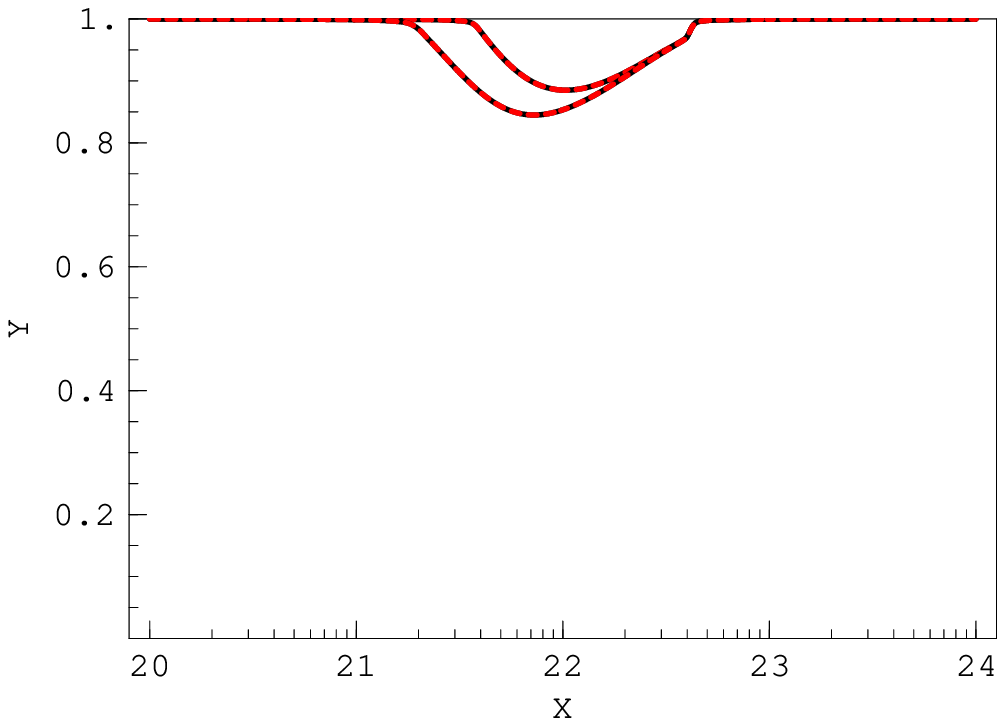,height=4.5cm,width=6cm,angle=0} \\[3ex]
 \psfig{figure=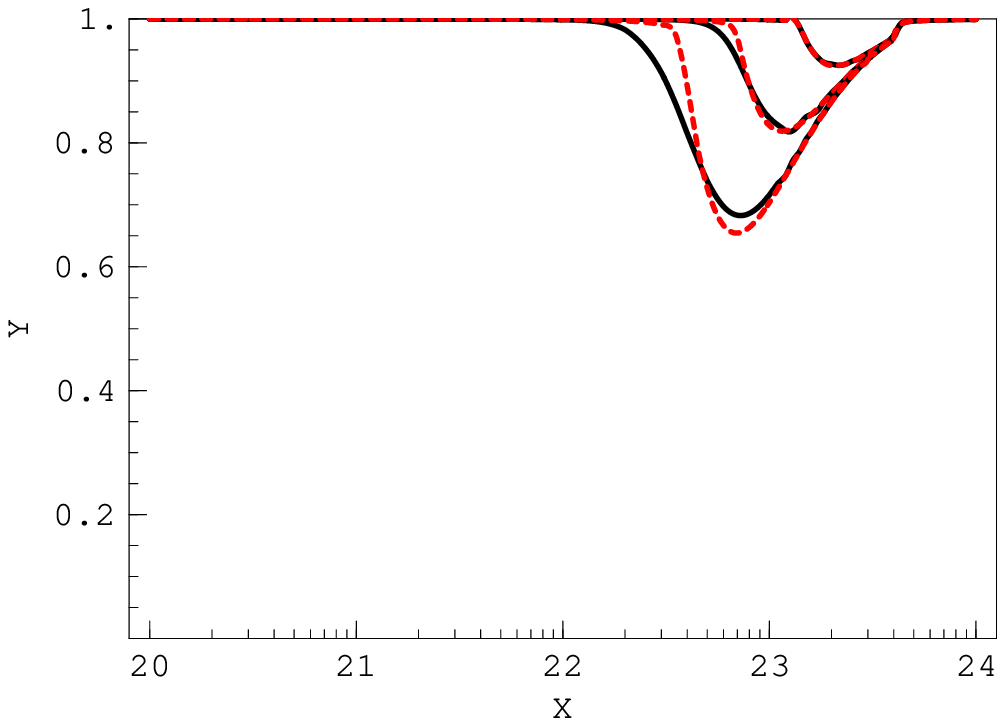,height=4.5cm,width=6cm,angle=0}
 \hfill
 \psfig{figure=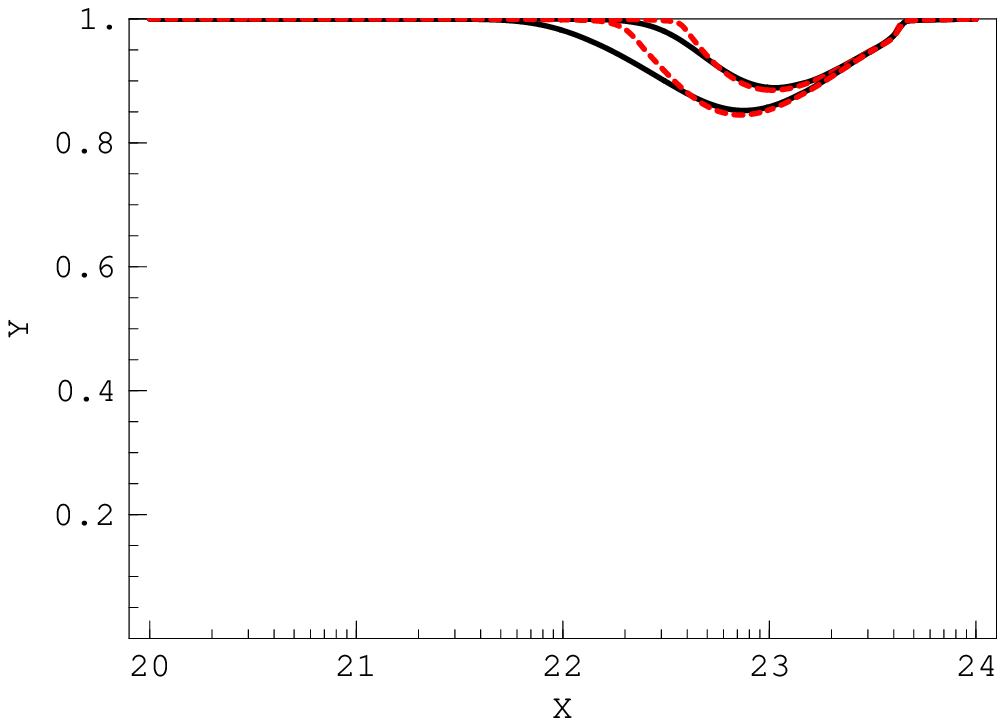,height=4.5cm,width=6
cm,angle=0}   \\[3ex]
 \psfig{figure=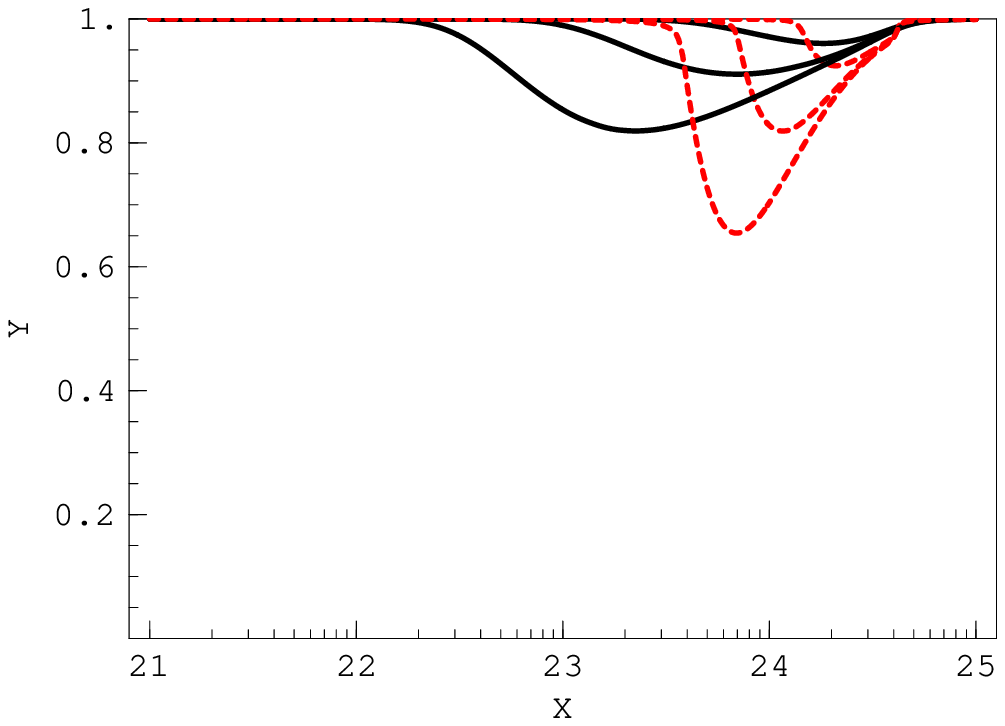,height=4.5cm,width=6cm,angle=0}
 \hfill
 \psfig{figure=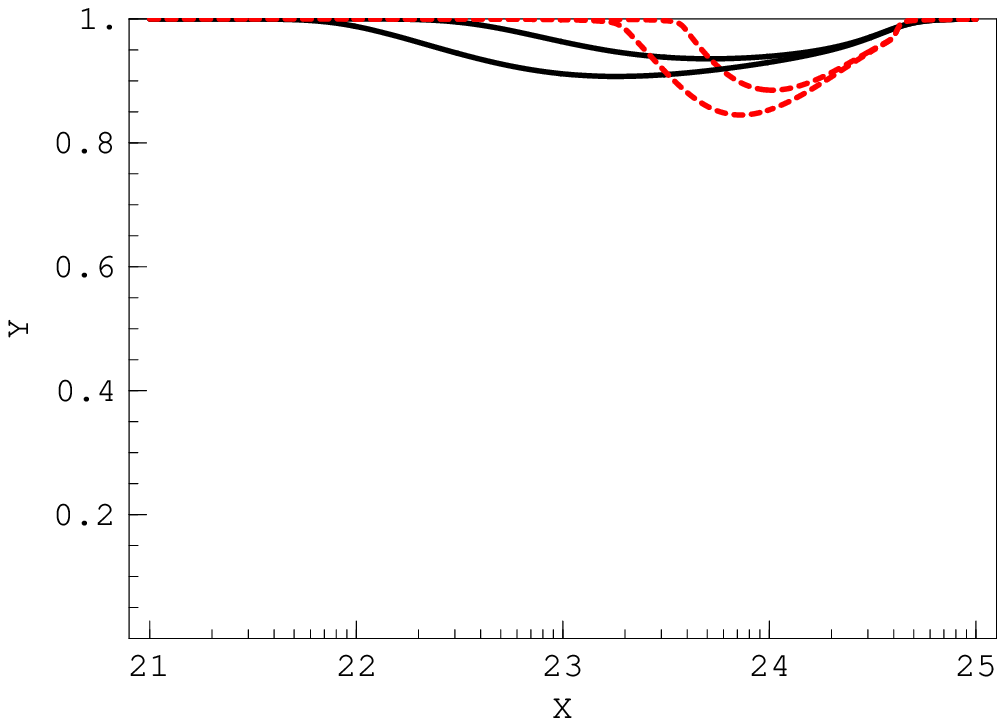,height=4.5cm,width=6
cm,angle=0}   \\[3ex]
 \psfig{figure=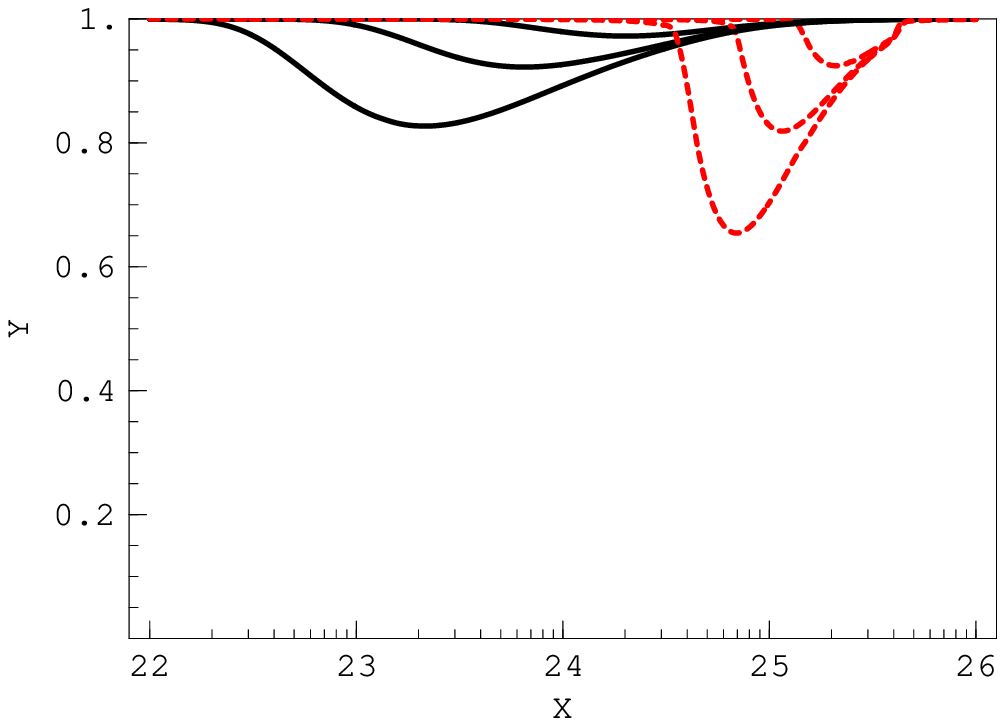,height=4.5cm,width=6cm,angle=0}
 \hfill
 \psfig{figure=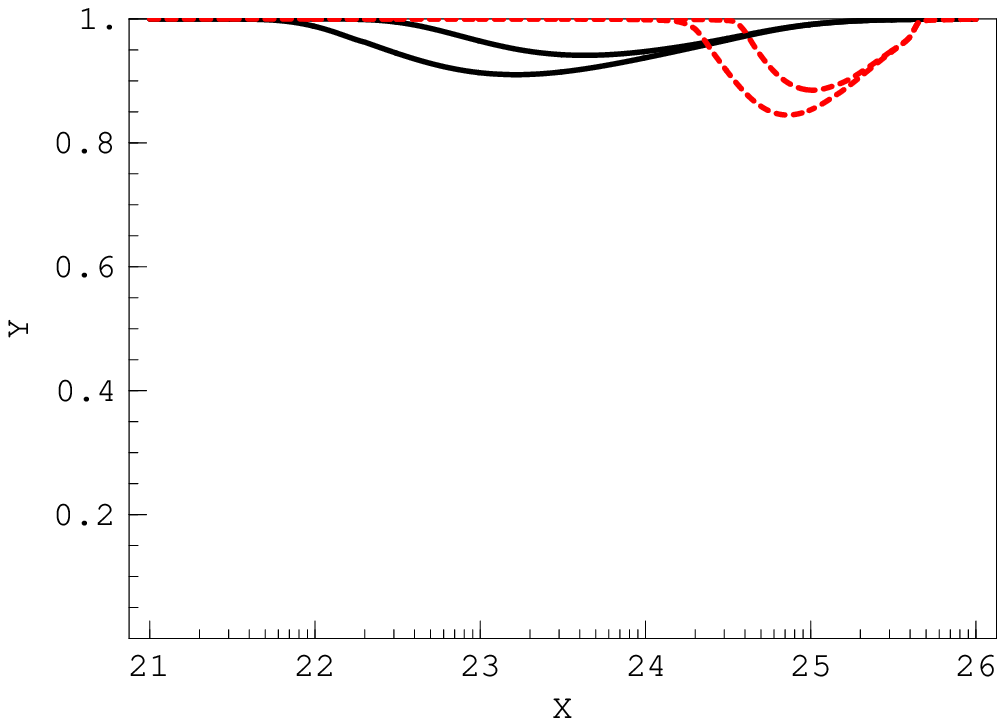,height=4.5cm,width=6
cm,angle=0}
 \caption{\footnotesize{UHE neutrinos fluxes in presence of damping, $\mathcal{F}_\nu$, normalized 
to the corresponding flux in absence of interactions, $\mathcal{F}_{\nu 0}$, for a neutrino mass 
$m_\nu = 10^{-1}, 10^{-2}, 10^{-3},10^{-4}$ eV (from top to bottom). The black (continued) curves 
are for the exact expression using eq.~(\ref{eq:sigmanubar}) while the dotted, red curves are for 
the approximation of neutrinos at rest, eq.~(\ref{eq:gammaRoulet}). The left column is for 
$n-\alpha = 2$ and $z_\mathrm{max} = 2$, $5$, $10$ (from top to bottom in each plot), while the 
right column is for $n-\alpha = 0$ and $z_\mathrm{max} = 10$, $20$.}} \label{fig:fluxes}
 \end{center}
 \end{figure}

The flux of UHE neutrinos at Earth depends on one hand on the mechanisms at work in the sources, 
which determine the injection spectrum of the UHE neutrinos, and on the other hand on the spatial 
and temporal distribution of the sources themselves. We follow here
the approach of~\cite{ringwaldnew} and express the \UHEnu flux in
function of the present neutrino 
energy $K_0$ as:

\begin{equation}
\label{eq:flux}
\mathcal{F}_\nu(K_0) = \frac{1}{4 \pi} \int_0^\infty \frac{\d z}{H(z)} \ P_\mathrm{T}(K_0,z) \ 
\mathcal{L}_\nu(K_0,z),
\end{equation}
 where $\mathcal{L}_\nu(K_0,z)$ is the neutrino source emissivity distribution, given in terms of 
the redshift $z$ and the present energy of the \UHEnu. In the hypothesis of identical injection 
spectra for all sources, one can factorize the dependence in the redshift and write
\begin{equation}
\mathcal{L}_\nu(K_0,z) = \eta(z)\ J_\nu(K_0),
\end{equation}
where $\eta(z)$ describes the distribution of the sources in the Universe, and  $J_\nu(K_0)$ gives 
the number of neutrinos emitted per unit of energy by each of these sources. We use the following, 
standard ansatz~\cite{ringwaldnew,sigl,semikoz}:
\begin{eqnarray}
\eta(z)&=&\eta_0 \ (1+z)^n\ \theta(z-z_\mathrm{min})\ \theta(z_\mathrm{max}-z); \label{eq:etaz}\\
J_\nu(K)&=& j_\nu \ K^{-\alpha}\ \theta(K_\mathrm{max}-K). \label{eq:jnu}
\end{eqnarray}
Eq.~(\ref{eq:etaz}) is suitable for an approximate description of UHE neutrino sources 
distribution in models ranging from astrophysical acceleration sites ("bottom-up" mechanisms, for 
which we can take $n\simeq 4$ and $z_\mathrm{max} \leq 10$) to exotic, non-accelerator sources 
(which have $n\simeq 1$ to $2$ and may extend to a much larger $z_\mathrm{max}$). In both cases, 
we take the lower bound for the source distribution to be $z_\mathrm{min} = 0$. The spectral index 
$\alpha$ typically ranges between 1 and 2, depending on the production mechanism considered; we 
assume here that it is constant over the range of energies examined and do not consider the 
possibility of broken power-law spectra above the GZK energy. We also suppose that $K_\mathrm{max} 
> K_\mathrm{res}(1+z)$ in all our analysis. Under these assumptions, as pointed out 
in~\cite{ringwaldnew}, the only dependence on the spectral indexes $\alpha$ and $n$ enters through 
a difference $n-\alpha$. We consider here, for the purpose of illustrating our results, two 
distinct situations: $n-\alpha = 2$ which could describe the UHE neutrino flux expected from an 
astrophysical, bottom-up-type source, and $n-\alpha = 0$ which would rather be associated to UHE 
neutrino fluxes produced in top-down processes. Results are presented in fig.~\ref{fig:fluxes}, 
which displays the \UHEnu flux, eq.~(\ref{eq:flux}), as a function of the present energy $K_0$ of 
the \UHEnu, after normalization to the flux in the absence of absorption effects, 
$\mathcal{F}_{\nu 0}$ (obtained by replacing $P_\mathrm{T}(K_0,z)=1$ in eq.~(\ref{eq:flux})). The 
first column corresponds to $n-\alpha=2$ and  for each value of $m_\nu$ we show several curves 
corresponding to different redshift limits for the source population, $z_\mathrm{max}=2$, $5$, 
$10$, while the second column corresponds to $n-\alpha = 0$ and redshift limits $z_\mathrm{max} = 
10$, $20$.

In that range of parameters, one can see that thermal effects do not significantly affect the 
shape of the absorption dip in the UHE neutrino flux as long as $m_\nu/T_{\nu} \gtrsim 10^2$. In 
particular, the endpoint of the dip
at high energies, corresponding to $K_\mathrm{res}$, is well-defined and can be used to estimate 
the absolute value of $m_\nu$, provided the absorption dip is not too shallow nor too
narrow to be resolved experimentally.
The global shape of the dip clearly depends on the value of
$n-\alpha$. Assuming that the
the spectral index $\alpha$ can be determined from the measurements of the \UHEnu spectrum in the 
range of energies which is not affected by the absorption, one could then obtain information on 
$n$ and therefore on the development of the source
population. From the
endpoint of the absorption dip at low energies, which corresponds to the resonance energy of the 
neutrinos emitted at the largest redshift, $K_0 (1+z_\mathrm{max})$, one
can also estimate the epoch at which the \UHEnu sources appeared.

For smaller values of  the ratio $m_\nu/T_{\nu}$, the situation
is significantly complicated by the  thermal motion of relic neutrinos. As a
result of the broadening of the transmission probability, the dips
get shallower and can extend over several orders of magnitude in
energy, depending on the maximum redshift chosen for the source
distribution. This might complicate their observation, especially
at small redshifts. From the figure one sees indeed that for
$z_\mathrm{max} \leq 2$, the flux won't be depleted more than a 5~\%.
Results will also be more difficult to interpret in terms of a
prediction for the neutrino mass and the maximum redshift for the
population of sources, since the position of the end points of the
absorption dip is not so clearly defined anymore. On the other
hand, maximum absorption is now achieved
at lower energies $K_0$. The shift is significant, even more than an order of magnitude for very 
small masses, $m_\nu \approx 10^{-4}$ eV. In view of the
inverse-power-law form of the \UHEnu energy spectrum, this could
significantly help to improve the detection potential, even though
the energy range $10^{23} - 10^{24}$ eV,
is currently beyond the reach of the majority of \UHEnu experiments
planned (see for example~\cite{ringwaldnew}).

\subsection{Absorption lines due to relic neutrino clustering}

The possibility that massive relic neutrinos cluster onto dark matter halos has been intensively
studied, in particular in relation with the generation of the UHE cosmic rays through the Z-burst 
mechanism~\cite{weiler99,fargion99,waxman98}. Recent works have presented revised estimations of 
the density profiles and typical spatial extension of the neutrino 
clusters~\cite{singh03,ringwald04}, giving overdensities of the order of $10-10^4$ with
respect to the present background density. They can extend on scales $L\sim 0.01 -1$ Mpc, 
depending on the neutrino mass, the mass of the attracting halo, and its velocity dispersion 
(which is typically of the order of 200 km/s for a galaxy and 1000 km/s for a galaxy cluster). 
This last parameter also constrains the epoch at which clustering can start, since neutrinos 
cannot be efficiently trapped as long as their mean velocity,
\begin{equation}
\left< v_\nu \right> \simeq 1.6\ 10^2\ (1+z)\ (\frac{\mathrm{eV}}{\mathrm{m}})\ \mathrm{km}\ 
\mathrm{s}^{-1},
\end{equation}
 is larger than the velocity dispersion of the attracting galaxy or galaxy 
cluster~\cite{ringwald04}. For neutrinos with masses $\lesssim 1$ eV, clustering will thus take 
place at very small redshifts and we can safely ignore the effect of  the expansion of the 
Universe in
this analysis. Other important limiting factors to the clustering of neutrinos on large scales are 
Pauli blocking and the limit on the maximum phase-space density, as
described in~\cite{Tremaine79}. They actually imply that only
neutrinos with mass $m_\nu \gtrsim 1$ eV will 
efficiently cluster on galactic halos, on typical scales $L_\mathrm{G} \sim 50 $ kpc, while 
neutrinos with mass $m_\nu \gtrsim 0.1$ eV can cluster on scales as big as $L_\mathrm{C} \sim 1$ 
Mpc in halos associated to (super-)clusters of galaxies~\cite{weiler99,singh03,ringwald04}.

The effect of neutrino clustering is limited to small scales and the overdensities are not large 
enough to have a significant incidence on the damping of \UHEnu  travelling on cosmological 
distances, except maybe in the case of galaxy
superclusters like Virgo~\cite{ringwald05}.

To compute the absorption by clustered neutrinos, one has to
substitute a
suitable distribution function $f^\mathrm{cl}(P)$ inside the cluster into
eq.~\ref{eq:gammasigma}.
To make a simple estimation of the effect, 
we assume that \(f^\mathrm{cl}(P)\) does not depend on the
position, \textit{i.e.}, we take a constant neutrino
overdensity. Outside of the cluster, the neutrino
density is that of the \CnuB. For \(f^\mathrm{cl}(P)\), we make the ansatz of a
modified Fermi-Dirac distribution
\begin{equation}
  \label{eq:f-ansatz}
  f^\mathrm{cl}(P) = \frac{1}{2}\, \frac{e^{-\Phi/T_\nu}+1}{e^{(P-\Phi)/T_\nu} + 1}.
\end{equation}
This distribution parametrises reasonably well the distribution functions presented 
by~\cite{ringwald04} in terms of a single parameter \(\Phi\), keeping
the temperature \(T_\nu\) unchanged from the \CnuB. The neutrino density
which corresponds to eq.~\ref{eq:f-ansatz} is
\begin{equation}
  \label{eq:cluster-n}
  n_\nu^\mathrm{cl} = -\frac{T_\nu^3}{2\pi^2} (1+e^{-\Phi/T_\nu})\,
  \mathrm{Li}_3(-e^{\Phi/T_\nu}),
\end{equation}
where \(\mathrm{Li}_3(x)\) is the trilogarithm function. For a given
overdensity factor \(N^\mathrm{cl}\) we solve $n_\nu^\mathrm{cl} = N^\mathrm{cl} \,n_{\nu 0}$ 
numerically for
\(\Phi\).

The clusters we consider in the following have an
overdensity factor $N^\mathrm{cl}$ between 10 and $10^4$ according to the 
results mentioned earlier, 
and a spatial extension $L_\nu^\mathrm{cl}$. We study the case where the cluster
is located between the source of the \UHEnu and the observer and
compute the transmission probability for neutrino masses of
\(1\,\mathrm{eV}\) and \(0.1\,\mathrm{eV}\).

\begin{figure}
 \begin{center}
 \psfrag{23}[c]{\tiny \phantom{n} \raisebox{0.1cm}{$10^{23}$}}
 \psfrag{22}[c]{\tiny \phantom{n} \raisebox{0.1cm}{$10^{22}$}}
 \psfrag{21}[c]{\tiny \phantom{n} \raisebox{0.1cm}{$10^{21}$}}
 \psfrag{20}[c]{\tiny \phantom{n} \raisebox{0.1cm}{$10^{20}$}}
 \psfrag{1.}[c]{\tiny \phantom{ni} \raisebox{0.1cm}{$1$}}
 \psfrag{0.8}[c]{\tiny \phantom{ni} \raisebox{0.1cm}{$0.8$}}
 \psfrag{0.6}[c]{\tiny \phantom{ni} \raisebox{0.1cm}{$0.6$}}
 \psfrag{0.4}[c]{\tiny \phantom{ni} \raisebox{0.1cm}{$0.4$}}
 \psfrag{0.2}[c]{\tiny \phantom{ni} \raisebox{0.1cm}{$0.2$}}
 \psfrag{0.}[c]{\tiny \phantom{ni} \raisebox{0.1cm}{$0$}}
 \psfrag{X}[c]{\tiny \raisebox{-0.5cm}{$K_0 [\mathrm{eV}]$}}
 \psfrag{Y}[c]{\tiny {$P_\mathrm{T}$}}

 \psfig{figure=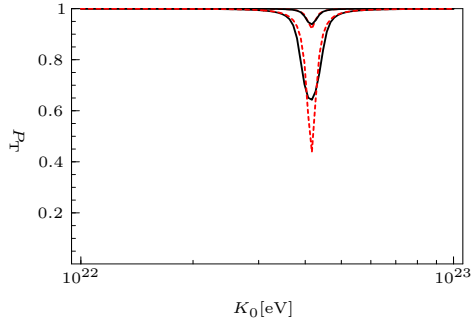,height=4cm,width=6cm,angle=0}
 \hfill
 \caption{\footnotesize{Transmission probability for a cluster of extension 1 Mpc, made of 
neutrinos of mass 0.1 eV, with a constant neutrino density $n^\mathrm{cl}_{\nu} = 10^3\ n_{0\nu}$ 
and $n^\mathrm{cl}_{\nu} = 10^4\ n_{0\nu}$.   }}
\label{fig:cluster}
 \end{center}
 \end{figure}

As expected, the effect of the thermal motion of the relic neutrinos is in general negligible or 
small due  to the relatively small overdensities achievable. We have to saturate the bounds on the 
parameters to get a significant effect, as shown in fig.~\ref{fig:cluster}, which displays the 
transmission probability with and without thermal effects, for a cluster of relic neutrinos with 
mass $m_\nu = 0.1$ eV and extension 1 Mpc. For a maximal
overdensity factor $N^\mathrm{cl}=10^4$ the thermal motion reduces the maximum
absorption probability across the cluster from $\approx 55 \%$
to $\approx 35 \%$, contributing to reducing its effect respect
to the non-clustered \CnuB absorption probability shown in
fig.~\ref{fig:transmission}.

\section{Conclusions}
\label{sec:conclusions}
Using the formalism of finite-temperature field theory we have calculated the damping factor of an 
\UHEnu propagating through the \CnuB including the effects of the thermal motion of the relic 
neutrinos in a systematic way. This allowed us to generalize the expressions for the transmission 
probability $P_\mathrm{T}$ that are commonly used in the
literature.

From the exploration of the parameter space allowed by cosmological and astrophysical constraints 
as well as by current limits on the neutrino mass, we see that thermal effects significantly 
affect the shape and position of the absorption dips in a realistic \UHEnu flux as soon as the 
ratio between the neutrino mass and the \CnuB temperature goes below $\approx 10^2$, {\it i.e.}, 
well before the relic neutrinos become relativistic. As expected, the effect essentially consists 
in smearing out the dip and shifting its minimum value to lower energies. This will complicate the 
observation of the dip in a real experiment measuring neutrino fluxes, especially if the \UHEnu 
source population is concentrated at small redshifts, producing rather shallow and extended dips. 
The shift of the absorption dip to lower energies, where neutrino fluxes are expected to be 
higher, increases in principle the potential of detection of the effect respect to the case of a 
neutrino at rest with the same mass. Still, the situation could become more intricate if the 
pattern of neutrino mass eigenstates is such that their combined effect results in a superposition 
of dips with different depths and extensions.

In the context of neutrino mass spectroscopy, we see from the examples that thermal effects do not 
affect the determination of the endpoint of the absorption dip, and hence the possibility of 
extracting information on the absolute neutrino mass, as long as $m_\nu \gtrsim 0.01$ eV, which is 
verified by at least one neutrino in the currently favoured mass
schemes~\cite{bell}. As the absorption dips get broader and shallower, the prospects for 
determining efficiently the resonance energy get worse as the endpoint of the absorption dip is no 
longer sharply defined.

Finally, as another application of our formalism, we have also investigated the transmission 
probability for UHE neutrinos propagating through a relic neutrino cluster. In the standard 
context of neutrino clustering around galaxies or galaxy clusters, we found that the thermal 
motion can further reduce the absorption effect of the cluster.
This absorption is small compared to the absorption of
\UHEnu in the \CnuB for neutrinos travelling cosmological distances,
since clustering only occurs at small redshifts.

\section*{Acknowledgments}

We would like to acknowledge support by CONACyT under grants 34868-E
and 46999-F and by DGAPA-UNAM under grants PAPIIT IN116503, IN119405
and IN112105.

\end{document}